\documentclass[a4paper,fleqn,table]{cas-sc}
\usepackage{tabularx}
\usepackage{stfloats}
\usepackage{multirow}
\usepackage{amsmath}
\usepackage{mdframed}
\usepackage{multicol}
\usepackage{longtable}
\usepackage{lipsum}
\usepackage{graphicx}
\usepackage{framed}
\usepackage[numbers]{natbib}
\usepackage{nomencl}
\makenomenclature

\begin{document}

\title[mode = title]{Towards Cyber Security for Low-Carbon Transportation: Overview, Challenges and Future Directions}

\cortext[1]{Corresponding author}

\author[1]{Yue Cao}[type=editor, auid=000, bioid=1]

\author[1]{Sifan Li}

\author[1]{Chenchen Lv}

\author[1]{Di Wang}[orcid=0000-0003-0104-3136]
\cormark[1]
\ead{2020102210009@whu.edu.cn}

\author[2]{Hongjian Sun}

\author[3]{Jing Jiang}

\author[4]{Fanlin Meng}

\author[5]{Lexi Xu}

\author[5]{Xinzhou Cheng}

\address[1]{School of Cyber Science and Engineering, Wuhan University, Wuhan 430072, China}
\address[2]{Department of Engineering, Durham University, Durham, DH1 3LE, UK}
\address[3]{Department of Mathematics, Physics and Electrical Engineering, Northumbria University, Newcastle, NE1 8ST, UK}
\address[4]{Alliance Manchester Business School, The University of Manchester, Manchester, M15 6PB, UK}
\address[5]{China Unicom Research Institute, Beijing, 100048, China}

\begin{abstract}[S U M M A R Y]
In recent years, low-carbon transportation has become an indispensable part as sustainable development strategies of various countries, and plays a very important responsibility in promoting low-carbon cities.
However, the security of low-carbon transportation has been threatened from various ways.
For example, denial of service attacks pose a great threat to the electric vehicles and vehicle-to-grid networks.
To minimize these threats, several methods have been proposed to defense against them.
Yet, these methods are only for certain types of scenarios or attacks.
Therefore, this review addresses security aspect from holistic view, provides the overview, challenges and future directions of cyber security technologies in low-carbon transportation.
Firstly, based on the concept and importance of low-carbon transportation, this review positions the low-carbon transportation services. Then, with the perspective of network architecture and communication mode, this review classifies its typical attack risks.
The corresponding defense technologies and relevant security suggestions are further reviewed from perspective of data security, network management security and network application security.
Finally, in view of the long term development of low-carbon transportation, future research directions have been concerned.
\end{abstract}


\ExplSyntaxOn
\keys_set:nn { stm / mktitle } { nologo }
\ExplSyntaxOff

\maketitle

%
%


\section{Introduction}\label{Indro}
According to the latest report of the intergovernmental panel on climate change "Climate Change 2022: Mitigation of Climate  Change" \cite{IPCC}, the world is likely to suffer extreme climate impacts unless global greenhouse gas emissions peak by 2025.
To reduce carbon emissions, various countries have introduced relevant policies, such as low-carbon transportation, to alleviate carbon emissions.
The "low-carbon transportation" literally means choosing a way for transportation service with low energy consumption, low emission and low pollution, as the general trend of urban sustainable development.
Here, the "low-carbon" refers to reducing the emission of greenhouse gases (mainly CO$_2$).

The emergence of electric vehicles (EVs) smart charging and smart grid collaboration technology is helpful to accelerate the evolution of low-carbon transportation \cite{glitman2019role}.
The EV smart charging and smart grid collaboration technology can reduce the emission of vehicle pollutants by EV orderly charging and vehicle-to-grid (V2G) technology \cite{MWASILU2014501}.
Moreover, the EV charging and battery swapping mainly obtain electricity from smart grid.
The auxiliary charging includes vehicle-to-vehicle (V2V) charging and other charging methods.
As the intermediary between the EV and the smart grid, the charging pile can realize the bidirectional flow of electricity \cite{MWASILU2014501}.
It can supply electricity to EVs, and also recycle the rich electricity of EVs.
In recent years, since renewable energy sources (such as solar energy, tidal energy, wind energy, water energy, etc.) generation is environmental friendly, it has become a part of the electricity source of smart grid \cite{eltigani2015challenges}.
\begin{figure*}
	\centering
	\begin{mdframed}
		\begin{multicols}{2}
			\textbf{Abbreviation}
			\begin{description}
				\item[{\makebox[4em][l]{EV}}]  electric vehicle
				\item[{\makebox[4em][l]{V2X}}] vehicle-to-everything
				\item[{\makebox[4em][l]{V2V}}] vehicle-to-vehicle
				\item[{\makebox[4em][l]{V2G}}] vehicle-to-grid
				\item[{\makebox[4em][l]{V2I}}] vehicle-to-infrastructure
				\item[{\makebox[4em][l]{ICTs}}] information and communication technologies
				\item[{\makebox[4em][l]{DoS}}] denial of service
				\item[{\makebox[4em][l]{CAN}}] controller area network
				\item[{\makebox[4em][l]{LiDAR}}] light detection and ranging
				\item[{\makebox[4em][l]{ECU}}] electronic control unit
				\item[{\makebox[4em][l]{CS}}] charging station
                \item[{\makebox[4em][l]{MITM}}] man-in-the-middle
				\item[{\makebox[4em][l]{IoV}}] internet of vehicle
				\item[{\makebox[4em][l]{XOR}}] exclusive OR
				\item[{\makebox[4em][l]{RSU}}] road side unit
				\item[{\makebox[4em][l]{CP}}] charging pad			
				\item[{\makebox[4em][l]{HMAC}}] Hash-based message authentication code				
				\item[{\makebox[4em][l]{GPS}}] global positioning system				
				\item[{\makebox[4em][l]{OCPP}}] open charging point protocol				
				\item[{\makebox[4em][l]{IP}}] internet protocol				
				\item[{\makebox[4em][l]{DDoS}}] distributed denial of service				
				\item[{\makebox[4em][l]{EVCS}}] electric vehicle charging station				
				\item[{\makebox[4em][l]{SDN}}] software-defined networking				
				\item[{\makebox[4em][l]{VANETs}}] vehicular Ad-Hoc networks 				
				\item[{\makebox[4em][l]{RSSI}}] received signal strength indication				
				\item[{\makebox[4em][l]{IDS}}] intrusion detection system				
				\item[{\makebox[4em][l]{EC}}] edge computing			
			\end{description}
		\end{multicols}
	\end{mdframed}
\end{figure*}

Cyber security is of importance in the reliability and stability of low-carbon transportation system. 
Firstly, data security is the first line of defense to guarantee confidentiality, unforgeability, and non-repudiation. 
Secondly, to ensure the reliability, and the integrity and correctness of data exchanged within network system, there is necessity to implement network management security, including trust management, misbehavior detection, intrusion detection system (IDS), and firewall. Finally, personalized network applications lead to different security requirements, in which edge computing security and software-defined security should be paid attention to.

To this end, this review focuses on analyzing and studying the ecosystem operation of low carbon, with major focus on the cyber security of low-carbon transportation.
The target of this review is to unlock the importance and impact of cyber security in low-carbon transportation and promote the coordinated development of EV, transportation, energy, information and cyber security.
Combining with the latest blockchain, software-defined networking (SDN) and edge computing (EC) technologies, this review summarizes and discusses the methods of attacks defense such as authentication, encryption, trust management, misbehavior detection, IDS, and firewall.
Towards cyber security for low-carbon transportation, the main contribution of this review is as follows:
\begin{itemize}
	\item Based on the concept and importance of low-carbon transportation, this review positions the low-carbon transportation service.
	\item Review of cyber security challenges and their mitigation solutions.
	\item In view of the current development trend of low-carbon transportation, challenges and research directions have been discussed.
\end{itemize}

The rest of this review is organized as follows: Section \ref{s2} introduces application scenarios, digitalized technologies, security incidents and typical attacks on low-carbon transportation. In Section \ref{datasecurity}, \ref{networkman}, related defense technologies (such as authentication, encryption, trust management, misbehavior detection, IDS, firewall etc.) are listed. In Section \ref{networkapp}, network application security which includes SDN, EC security is introduced. Finally, challenges and future research directions are discussed in Section \ref{future}. 

\section{Background}\label{s2}
The integration of transportation, energy and information networks could promote the development of low-carbon transportation.
However, various application scenarios and information interaction in low-carbon transportation will expose its vulnerable attacks.
For attacks in application scenarios or communication modes, a security defense system is compulsory.
This section firstly lists typical application scenarios and components of low-carbon transportation.
Secondly, it analyzes typical information and communication technologies (ICTs) about E-Mobility, smart grid, in-vehicle and vehicle-to-everything (V2X) communication.
Finally, potential cyber risks and typical attacks on low-carbon transportation are given.

\subsection{Components of Low-Carbon Transportation}
\begin{figure*}[!htbp]
\centering
\includegraphics[scale=0.45]{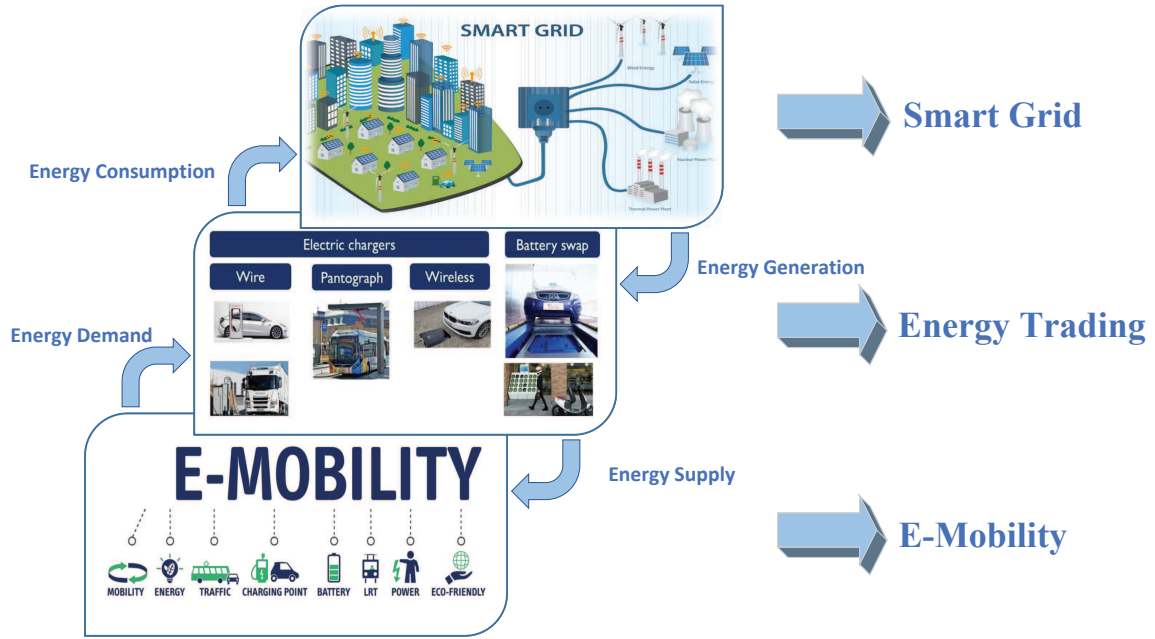}
\caption{Concept of Low-carbon Transportation}\label{all}
\end{figure*}

In Figure \ref{all}, as the energy consumers, EV could obtain energy from charging station (CS) or battery swap station.
The smart grid generates energy to supply CS, battery swap station, etc.
As an intermediary, CS and battery swap station process the energy demand from EV.

\subsubsection{EV Application Scenarios}
Due to the environmental advantages brought by EVs, the marketization of EVs is steadily advancing.
Following this global trend, traditional car companies, such as Mercedes-Benz and BMW, or EV company Tesla, have been planning transformation towards vehicle electrification. 
Since 2009, with the strong support of national subsidy policy, China has vigorously promoted the development of EVs.
As of 2019, the annual sales of EVs in China was 1.206 million, ranking the first for five consecutive years.
To realize the transition of EVs from policy-led to market-driven in China, the General Office of the State Council issued the "New Energy Automobile Industry Development Plan (2021-2035)" \cite{113} in November 2020.

\begin{figure*}[!htbp]
	\centering
	\includegraphics[scale=0.15]{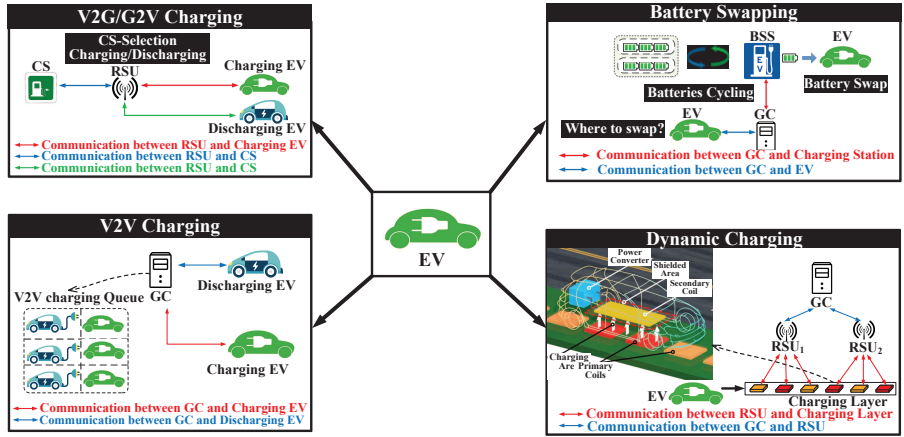}
	\caption{EV Application Scenarios}\label{fig1}
\end{figure*}

Figure \ref{fig1} shows typical EV application scenarios.
At present, the charging modes of EVs are divided into grid-to-vehicle (G2V) charging, V2G charging, V2V charging, and dynamic charging.
The G2V refers to the smart grid provides energy supply to the EV with insufficient electricity.
Conversely, the V2G means that the EV with sufficient electricity sells the excess electricity back to the smart grid, ensuring that the load balance of smart grid during peak electricity consumption.
Besides, the V2V charging refers that the EV with sufficient-electricity, is able to supply those with insufficient electricity in a peer-to-peer manner.
The dynamic charging is that the driving EV is charged by the charging pads (CPs) laid on the ground through electromagnetic induction. However, compared with the charging modes including G2V and V2G, the battery switching mode achieves a shorter service time, by replacing a depleted battery with a fully charged one.

\begin{figure}[!htbp]
	\centering
	\includegraphics[scale=0.4]{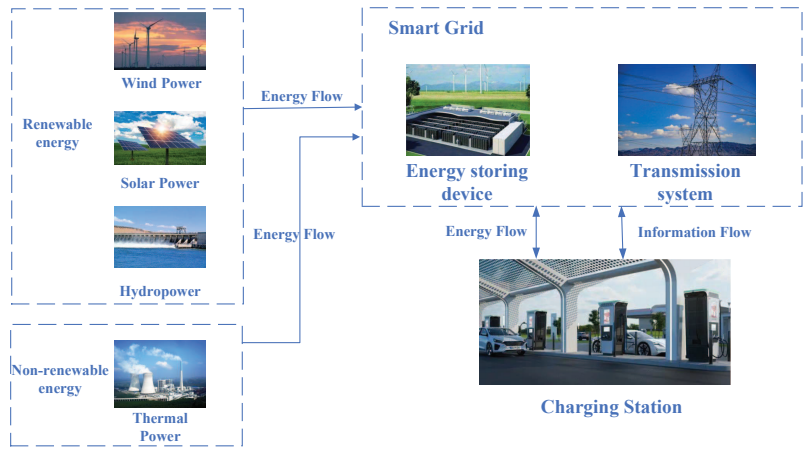}
	\caption{Smart Grid Application Scenarios}
	\label{SGscene}
\end{figure}

\subsubsection{Smart Grid Application Scenarios}
One of the core concepts of smart grid is compatibility, including centralized and distributed power generation infrastructures and access of various energy storage devices.
Distributed power refers to the generation and storage of various small-sized devices connected to the smart grid or the distribution system, known as distributed energy resources (DER).
The DER system is a decentralized, modular and more flexible technology, utilizing renewable energy including small-scale solar energy and wind energy.
Otherwise, conventional power stations (such as coal, gas and nuclear power plants, hydropower dams and large-scale solar power stations) are centralized power.
As shown in Figure \ref{SGscene}, small-scale solar power and wind power belong to distributed power, while thermal power belongs to centralized power.
Under the unified dispatching and control of smart grid system, the V2G technology is developed to realize the two-way exchange of information flow and energy flow.
Through the cooperation of intelligent dispatching power generation capacity, energy storage facilities and intermittent renewable energy generation, the load demand of users can be better implemented.

The 5G and artificial intelligence empowered smart grid will help improve energy transportation and utilization efficiency. 
For example, in 2020, the China Telecom announced that the first phase of Qingdao 5G smart grid project has been jointly developed with State Grid Qingdao Power Supply Company, China Telecom Qingdao Branch and Huawei, marking the formal completion of the largest 5G smart grid in China \cite{9310746}. 
Besides, DeepMind and the National Grid plan to add DeepMind's artificial intelligence technology to the UK's smart grid system \cite{DeepMind}. The project will deal with a large amount of information such as weather forecast and internet search, so as to develop a forecasting model of demand surge.

\subsubsection{V2X Communication Application Scenarios}
\begin{figure}[!htbp]
	\centering
	\includegraphics[scale=0.4]{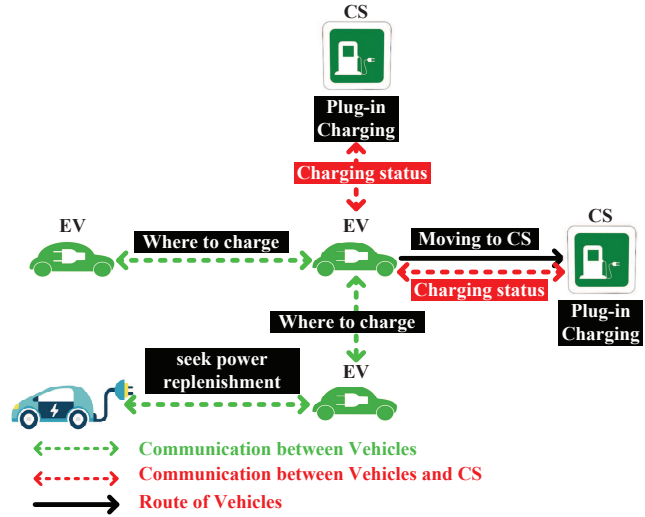}
	\caption{The Diagram of V2X Communication Scenario}
	\label{scene}
\end{figure}

The limited EV battery capacity and long trips in urban cities lead to frequent charging demands of EVs.
In general, CSs are typically deployed in places with high EVs density, e.g., shopping malls and parking lots.
However, in case CS has no sense of arrival of EVs with charging demand, a potential charging service congestion may occur.
Benefiting from V2X communication technology, it is possible to share such knowledge to optimally manage the charging process.
For example, as shown in Figure \ref{scene}, an EV seeking power replenishment obtains the charging status of CSs through V2X communication, then moving towards the CS with the earliest charging service availability.
In addition, by adopting V2X communication, EVs can also actively send requests to seek power replenishment from suitable EVs in a peer-to-peer manner.

At present, the traffic planning is of great importance for future city and its economic improvement.
The emergence of V2X can promote the development of traffic planning.
The application of V2X technology makes it possible to realize intelligent traffic management.
It can accurately classify vehicle types and rationally arrange vehicle routes according to the traffic jams in the urban areas, so it can improve traffic efficiency and bring additional productivity. 
For example, London is upgrading urban infrastructure by leading the way in 5G infrastructure.
The Smart Mobility Living Lab in London (SMLL) is deploying the world most advanced urban testbed.
The city will test vehicle-to-infrastructure (V2I) and V2V capabilities in a real world environment, by taking advantage of high-speed 5G. 
The city launched Sitraffic Fusion, a program designed to manage traffic through data obtained from connected vehicles.

\subsection{ICTs in Low-Carbon Transportation}
\subsubsection{ICTs in Smart Grid}
At present, the State Grid Corporation of China has built a power backbone communication network, based on optical fiber communication and various communication methods (such as microwave, carrier) coexisted.
The backbone communication network consists of transmission network, business network and supporting network.
In particular, the transmission network includes optical cable, optical communication, microwave communication, satellite communication, carrier communication, etc.

\subsubsection{Open Charging Point Protocol (OCPP)}
Despite the evolution of EVs has been paid extensively from both academia and industrial sides, factors such as imperfect charging facilities, non-standardization and non-uniform standards restrict the development of EV industry.
Under the joint efforts of policies and industries in the world, EVs and charging facilities are expanding rapidly. 
Therefore, it is necessary for charging standards and protocols to be standardized.
The open protocol allows the interconnection of CSs and back-end systems, without special information interfaces or gateways. The OCPP is a general-purpose open communication protocol that addresses the challenges posed by private charging networks.

\subsubsection{Charging Pile Communication}
The communication mode of charging pile is mainly classified into wired communication and wireless communication modes.
The wired communication mode mainly includes industrial serial bus, wired Ethernet, etc.
Here, industrial serial bus data transmission is more reliable, but with high complexity, low communication capacity, poor flexibility, high construction cost and poor scalability.
The wired Ethernet network has higher capacity and reliable data transmission, despite of complex wiring, poor flexibility, high construction cost and poor expansibility.


\subsubsection{V2X Communication Technologies}

The V2X enables all-round communication among vehicles, people, transportation infrastructures and cloud centers.
Here, V2X communications are mainly divided into four categories: V2V, V2I, vehicle-to-pedestrian (V2P), and vehicle-to-network/cloud (V2N/C).
Among them, V2V, V2I and V2P are with characteristics such as low delay, high reliability and dynamic connectivity, while V2N/C has slightly lower requirements on these characteristics.

\subsection{Security Incidents in Low-Carbon Transportation}
\subsubsection{Smart Grid}
Due to the upgrade of ICTs, the concept of smart grid has been introduced in past few years.
In the smart grid, fruitful information is exchanged and monitored to trigger the decision making of power system control. However, although the access network improves the efficiency of system operation, it will endanger the security of system because of its dependence on information flow.

Since communication links are vulnerable for cyber attacks and enemy infiltration, the smart grid is not immune to attacks by hackers, whether motivated by politics or profit, that have proliferated in recent years.
In 2015, 1.4 million residents of Kiev and western Ukraine suddenly found their homes with electricity cut-down, not because of a power shortage, but due to cyber attacks \cite{7752958}.
According to the Head of cyber espionage intelligence at iSight Partners, the cyber attack, known as BlackEnergy, ordained from a Russian hacking group using malware.
The Black Energy generates client programs for infected hosts, and command scripts for architecture in command and control (C\&C) servers.
Attackers can easily set up botnets by using hacker software. Attackers can issue simple instructions in the Command and Control server, then botnet victim hosts will uniformly execute their instructions.

In 2019, hackers used known vulnerabilities in Cisco's firewalls to launch a denial of service (DoS) attack on a Utah renewable power company \cite{hale2020utilities}. The incident affected California, Utah and Wyoming. 
The North American Electric Reliability Corporation (NERC) reported in September that the security breach affected the Web interface of firewalls used by the victims, and the attackers triggered DoS conditions on those devices, causing them to restart. 
This caused a breakdown in communication between the organization's control center and field equipment at its various sites.

In June 2020, a Brazilian power company, Light S. A, was hacked and ransomed for \$14 million. 
Security research at AppGate analyzed it as Sodinokibi ransomware \cite{he2020detection}. 
Sodinokibi can be used in ransomware as a service (RaaS) mode, it may be operated by threats associated with Pinchy Spider (the organization behind ransomware). The research also found that the software can enhance privileges by exploiting 32-bit and 64-bit vulnerabilities of CVE-2018-8453 in Win32k components. 
In addition, the ransomware family does not have a global decryptor, meaning that the attacker's private key is required to decrypt files.

\subsubsection{Vehicle System}
Francillon et al. \cite{francillon2011relay} elaborated threat of relay attacks against passive keyless entry and start systems.
The attacker amplified the signal from the key and car, so that both could receive and respond to radio frequency signals over a short distance.
By completing the challenge-response communication process, the attacker finally achieved the goal of unlocking vehicles.
In addition to relay attacks, Wouters et al. \cite{wouters2019fast} found outdated encryption algorithms and insufficient mutual authentication during keyless-to-vehicle authentication.
The Tencent security research team conducted security analysis on the electronic control units (ECUs) of various BMW models, and found $14$ general security vulnerabilities, involving in-vehicle infotainment systems, in-vehicle communication modules, and in-vehicle gateways \cite{keenexperimental}.
In addition, the team also identified security vulnerabilities for in-vehicle infotainment system and in-vehicle communication module of Mercedes-Benz \cite{keenbenz}.

Keuper et al. \cite{keuper2018connected} found that the vehicle could be connected to the Wi-Fi network through the in-vehicle Wi-Fi device.
By exploiting vulnerabilities for in-vehicle infotainment system, authors could send controller area network (CAN) messages to the CAN bus so as to control the center screen, speakers, microphones, etc.

Hackers exploited vulnerabilities in Tesla \cite{cyberattack} to remotely control the Model X via Wi-Fi or cellular networks, such as opening the doors or trunk, controlling the brakes, or controlling the car's radio to play music. 
The Keen security lab of Tencent \cite{Free} successfully injected malicious information into the CAN bus of Tesla Model S to remotely control parking and driving modes. After it submitted the vulnerability report to Tesla, Tesla introduced the code signing to defend against attacks.

Targeting autonomous vehicles, Cao et al. \cite{cao2019adversarial} proposed an optimization-based approach, namely light detection and ranging (LiDAR)-Adv, to generate realistic adversarial objects that can evade LiDAR-based detection systems under various conditions.
Through 3D-printed optimized adversarial objects, authors demonstrated that their work can consistently mislead LiDAR systems equipped on vehicles.
In addition, the research in Keen security lab analyzed the visual recognition system of autopilot ECU through static reverse engineering and dynamic debugging, and successfully implemented attacks \cite{tencent2019experimental}.
Through the generated adversarial sample stickers, the research successfully misled the automobile into the opposite lane, causing the wrong way.
Shen et al. \cite{shen2020drift} conducted safety research on global positioning system (GPS) for autonomous vehicles.
Research has shown that the multi-sensor fusion positioning scheme of an autonomous vehicle is attacked by means of "GPS spoofing", causing the vehicle to lose control.
This safety issue has sounded the alarm for manufacturers who have accelerated the commercialization of autonomous driving in recent years.

\subsection{Typical Attacks in Low-Carbon Transportation}
\begin{figure*}[!htbp]
	\centering
	\includegraphics[scale=0.60]{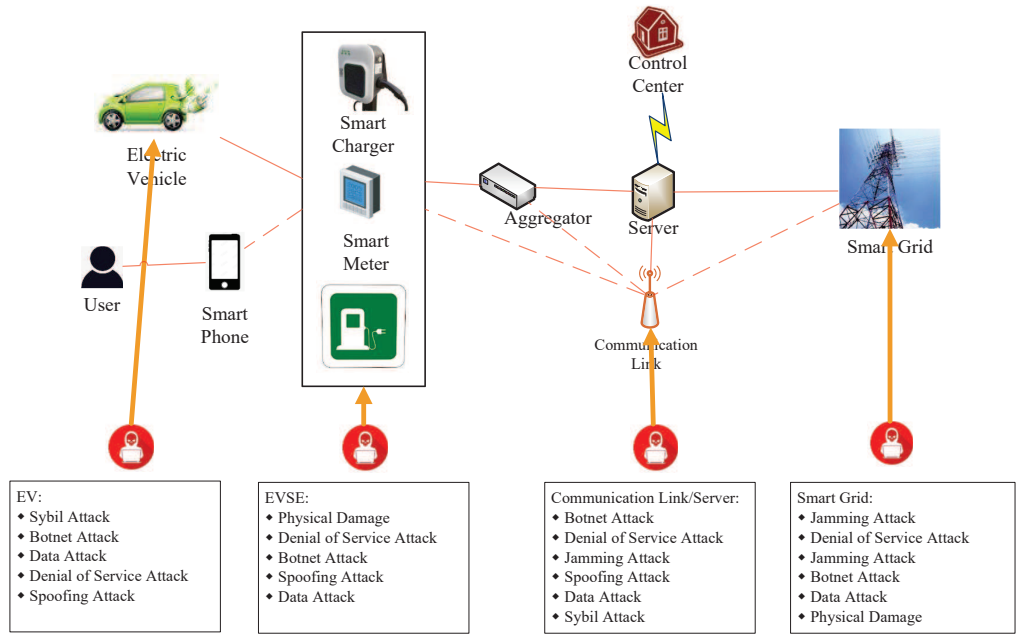}
	\caption{Six Typical Attacks Against EV, EVSE, Server, Communication link, and Smart grid}\label{v2gATTACK}
\end{figure*}

Cyber security is one of the major challenges that all countries in the world are facing at present. 
The promotion and development of low-carbon transportation can not be without the escort of cyber security.
The perfect cyber security protection and prevention system is the cornerstone of low-carbon city development.
However, the low-carbon transportation is facing many attacks and security threats.
The attacks of low-carbon transportation are mainly classified into jamming attack, spoofing attack, data dimension attack, DoS attack, botnet attack and sybil attack.
Note that, apart from typical cyber attacks, there is also vulnerability of physical attack by using violent means to attack EV, charging pile and other firmware etc. 
Here, regular safeguard and maintenance on those key infrastructures are needed.
Table \ref{Attack} classifies cyber attacks and summarizes defense solutions.
Figure \ref{v2gATTACK} shows six typical attacks and scenarios against EV, electric vehicle service equipment (EVSE), server, communication link, and smart grid.
\subsubsection{Jamming Attack}
The attacks apply wireless technology to interfere with the communication between EV and charging infrastructure.
Consequently, the EV may not receive the available status of CS when in need of charging.

\subsubsection{Spoofing Attack}
In order to disturb the network operation, the spoofing attack creates fake or spoofed identities to legitimately communicate with victim or send fake messages.
Attackers commonly used network spoofing methods mainly include address resolution protocol (ARP) spoofing, internet protocol (IP) spoofing, domain name spoofing, Web spoofing and email spoofing.

The attacker communicates with EV disguised as a CS, aggregator or charging pile.
If the EV is not authenticated, private data such as the identity, battery status and location of EV may be illegally released to attackers. Similarly, in case the charging infrastructure does not authenticate the EV, this may lead to malicious EV injecting abnormal data into charging infrastructure being attacked.

\subsubsection{Data Attack}
As an important channel across EVs, smart grid, charging pile, control center and infrastructure, the data transmission is vulnerable to tampering, injection, delay, eavesdropping, man-in-the-middle (MITM) attack etc.
For example, under the MITM attack, an attacker acts as an intermediary to receive, eavesdrop, forge information between EVs and charging infrastructures.
Besides, the attacker can inject false information into the EV battery status and sends to the CS, this misleads the decision of CSs.
When EV conducts energy transaction, attackers eavesdrop on the power information collected by CS and global controller (GC) to obtain sensitive information (such as EV's identity, location and user preferences).
Meanwhile, once sensitive data (such as the user's transaction password, amount and management password) is stolen and maliciously tampered with, it may cause damage to the smart grid and operation of EV.

\begin{table*}[!htbp]\small
	\caption{Comparisons of Attack Threats}\label{Attack}
	\centering
	
	\begin{tabularx}{\textwidth}{lXXXX}
		\toprule
		Types of attack & Attack Type & Attack target & Consequences & Solutions \\
		\midrule
		
		Jamming attack & Active & EV and charging infrastructure & EV unable to charge & Cloaking and acoustic\\
		
		Spoofing attack & Active & EV, EVCS & Private data leakage, EVCS infected & Digital signature\\
		
		Data dimension attack & Active\newline Passive(eavesdropping) & Data transmission & Data is stolen and corrupted & Authentication, Encryption\\
		
		DoS attack & Active & Charging pile, EVCS, EV, Smart Grid & EVCS can not work  & Authentication, IDS\\
		
		Botnet attack & Active & Charging pile, EVCS, EV, Smart Grid &  Loss of control & Authentication, IDS\\
		
		Sybil attack & Active & Global Controller & Network congestion and energy shortage & Authentication, Trust assessment test\\
		\bottomrule
	\end{tabularx}
\end{table*}

\subsubsection{DoS Attack}
The goal of a DoS attack is to prevent the target machine from providing service.
DoS is one of attacks that are difficult to solve and defend, where distributed denial of service (DDoS) is the most harmful one.
The difference between DDoS and DoS is that DDoS is a kind of distributed DoS.
It combines multiple computers to launch DDoS attacks on one or more targets, thus increasing the harm of DoS attacks by times.

DDoS attacks are created by hacking into smart grid devices, to gain control and directing them to launch targeted DoS attacks.
Charging piles and CSs are vulnerable to DoS attacks, mainly for the following reasons.
First, many charging piles or CSs are connected devices, and these smart grid devices often do not have password or password is relatively simple.
Second, the large number of smart grid devices are also easily facilitated by attackers to maximize the impact of DDoS.
Third, malicious smart grid devices are hard to detect even if devices are hacked.

Schneider Electric has announced that it has patched several new vulnerabilities in its EVlink CSs \cite{Schneider-Electric}. 
These include cross-site request forgery (CSRF) and cross-site scripting (XSS) vulnerabilities, which can be exploited to perform operations on behalf of legitimate users.
Schneider Electric warned that failure to act on the EVlink vulnerabilities, could lead to tampering and leaks of CS settings status, accounts, etc.
Such tampering can lead to DoS attacks, resulting in unauthorized use of CSs, interruption of service, failure to send changes about charging data records, disclosure of regulatory systems and CS configurations.
An attacker can exploit these vulnerabilities to control EV charging station management system (EVCSMS) and lock down the underlying EV charging station (EVCS) or disable specific features in its configuration, thereby denying physical and virtual access to legitimate clients.

\subsubsection{Botnet Attack}
Attackers can launch a variety of cyber attacks by infecting large numbers of botnets and hosts.
For example, an attacker can launch DoS attacks or internet probing/reconnaissance activities to a target.
To achieve this, an attacker can leverage server-side request forgery (SSRF) vulnerabilities to use the compromised EVCS as proxies.
Consequently, SSRF vulnerabilities force them to redirect requests towards internal/external endpoints, and perform lateral movement on the network as well as scan third-parties.
The infected EVCS could potentially be used to conduct cyber attacks on integrated infrastructure such as smart grid.
Attackers can also leverage a large number of infected EVCSMSs to initiate simultaneous charging operations, or reverse current flow back to the smart grid by increasing the discharge supply.
Both of these attacks destabilize the smart grid, by causing sudden increases in power load to disrupt the power load balance of smart grid, leading to cascading failures.

\subsubsection{Sybil Attack}
Attackers manipulate their identities and send false information to GC, leading to abnormal decision making of GC and failure of charging service.
Active and passive attacks are classified depending on the attack behavior.
Here, active attacks mean that the attackers will tamper, forge or discard data packets.
In contrast, passive attacks only eavesdrop on data or obtain the frequency of data transmission.

Internal and external attacks are classified according to the role of attackers.
Here, internal attacks refer to the authenticated users to launch the attacks.
In contrast, external attacks refer to the unauthenticated users to launch the attacks.

\begin{figure*}[!htbp]
	\centering
	\includegraphics[scale=0.8]{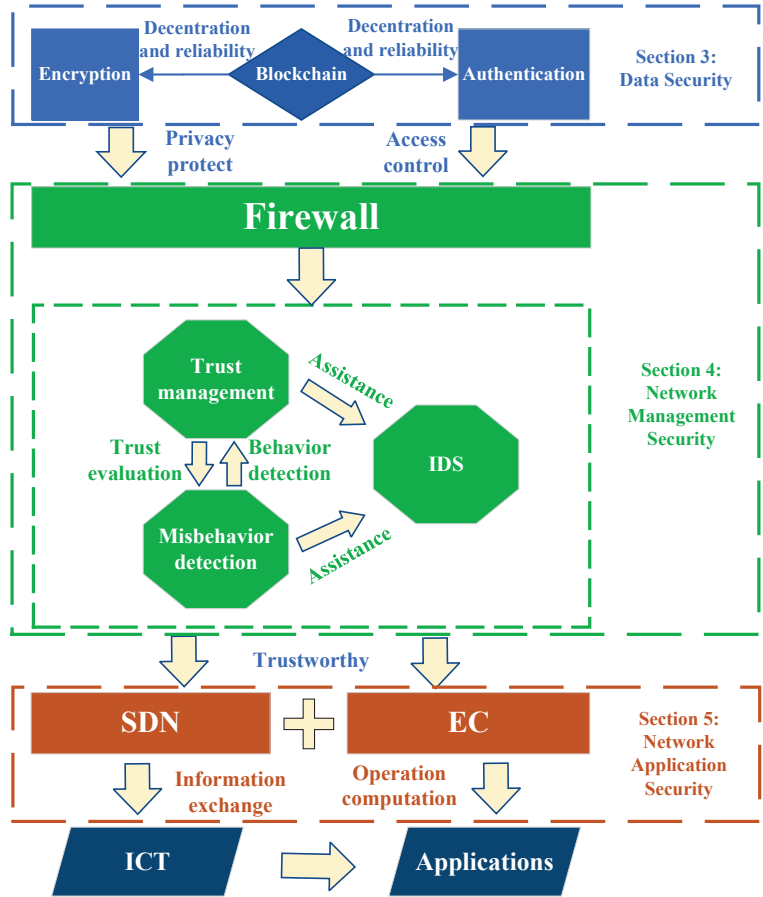}
	\caption{The Relationship Among Defense Technologies}\label{defense}
\end{figure*}

\subsection{Taxonomy of Defense Technologies}
Given typical attacks been introduced and investigated in lien with the use cases under low-carbon transportation system, dedicated defense technologies are of great importance. 
Here, depending on how those defense technologies perform within the low-carbon transportation system, we classify those into data security, network management security and network application security. Details of these can be referred at following sections, and there relations can be referred to Figure \ref{defense}.


\section{Data Security}\label{datasecurity}
The data security guarantees the confidentiality, unforgeability, and non-repudiation. In this section, literature on current defense technologies (i.e. encryption and authentication) are reviewed. 
To enhance data security, decentralization and traceability are essential, thus blockchain as an emerging technology is concerned. 
Here, the applications of blockchain in smart grid and vehicular Ad-hoc networks (VANETs) are also covered.

\subsection{Blockchain}
As the advent of bitcoin, blockchain began to rise as a distributed ledger \cite{46}. 
Cryptography and consensus ensure data security and data consistency, respectively. 
Moreover, chain structure traces the source and accountability. 
As tamper proof code, smart contract realizes the application of blockchain, where peer-to-peer network is applied to data dissemination and verification.

Due to its security, such as tamper resistance and consistency, blockchain has attracted much attentions and widely applied in smart grid \cite{47,48,49,50,51,52} and VANETs \cite{53,54,55,56,57,58}. The application of blockchain in smart grid mainly includes energy trading \cite{47,48}, demand response \cite{49,50}, and payment \cite{51,52}. The application of blockchain in VANETs mainly includes data sharing \cite{53,54}, ride sharing \cite{55,56}, and EC \cite{57,58}. Table \ref{APPSmartG} shows the application of blockchain in smart grid and VANETs.

\begin{table*}[!htbp]\small
	\caption{The Applications of Blockchain in Smart Grid and VANETs} \label{APPSmartG}
	\centering
	\begin{tabularx}{\textwidth}{cXX}
		\toprule
		Application Field & Application Scenario & References\\
		\midrule
		Smart Grid &
		\begin{itemize}
			\setlength{\itemsep}{0pt}			
			\setlength{\parsep}{0pt}
			\setlength{\parskip}{0pt}
			\item Energy trading
			\item Demand response
			\item Payment
		\end{itemize} &
		\begin{itemize}
			\setlength{\itemsep}{0pt}			
			\setlength{\parsep}{0pt}
			\setlength{\parskip}{0pt}
			\item \cite{47,48}
			\item \cite{49,50}
			\item \cite{51,52}
		\end{itemize}\\	
		VANETs &
		\begin{itemize}
			\setlength{\itemsep}{0pt}			
			\setlength{\parsep}{0pt}
			\setlength{\parskip}{0pt}
			\item Data sharing
			\item Ride sharing
			\item EC
		\end{itemize} &
		\begin{itemize}
			\setlength{\itemsep}{0pt}			
			\setlength{\parsep}{0pt}
			\setlength{\parskip}{0pt}
			\item \cite{53,54}
			\item \cite{55,56}
			\item \cite{57,58}
		\end{itemize}\\
		\bottomrule
	\end{tabularx}
\end{table*}

Li et al. \cite{47} proposed a consortium blockchain-based secure energy trading, and then utilized reputation mechanism to improve the speed of transaction. 
Finally, the stackelberg game is applied to optimize pricing. 
Although differential privacy can ensure data privacy, the introduction of noise affects the correctness and integrity of data. 
In response to these challenges, Gai et al. \cite{48} introduced noise into the neighboring energy trading system to prevent attackers from inferring the trading distribution trend.

Zhou et al. \cite{49} proposed an incentive-compatible power demand response scheme, by applying consortium blockchain to realize the creation, transmission and verification of transactions. 
Then, convex programming is applied to optimize social welfare. 
To ensure the load balance and stability of smart grid, Jindal et al. \cite{50} applied blockchain to initialize energy trading and uplink storage, in which energy trading includes responding to the energy demand of residential, commercial, and industrial areas. 
Moreover, this work has been applied to California to prove its effectiveness.

Gao et al. \cite{51} proposed a privacy protected payment scheme for V2G network, which requires vehicles to register and trade on the blockchain. 
The scheme is deployed in Hyperledger to verify its effectiveness. 
Dimitriou and Mohammed \cite{52} proposed a bitcoin-based payment scheme to encourage users to share power data to suppliers, reducing the cost of users.

To choose data providers, Gawas et al. \cite{53} proposed a reputation evaluation mechanism for blockchain combined with machine learning (ML). 
This scheme solves the sharp increase of block data and prevents attackers from damaging consensus. Facing that roadside units may be attacked in data sharing, Cui et al. \cite{54} proposed a data sharing and enhanced delegated proof-of stake without road side units (RSUs) assistance, to ensure the robustness of data and miners.

Li et al. \cite{55} proposed a ride sharing service with privacy protection, by applying private proximity test and private blockchain to realize proximity matching and data audit respectively. 
Baza et al. \cite{56} utilized blockchain to match appropriate drivers for riders. 
Moreover, zero knowledge proof and time lock are applied to verify pick up point and deposit respectively.

To address limited storage of edge nodes, Lu et al. \cite{57} proposed a flexible blockchain storage structure by rewriting the data without affecting the integrity of original data. 
Along with the increase of the number of nodes, the performance advantage of this scheme is more obvious. 
Yuan et al. \cite{58} leveraged the blockchain to provide efficient storage, reputation assessment, and rewards for the collaborative edge. 
This scheme promotes secure, reliable, and efficient completion of task collaboration and offloading.

\subsection{Authentication}
\subsubsection{Authentication Protocols for V2G Charging}
Table \ref{RWV2G} and Table \ref{RWV2G2} review authentication protocols in V2G charging, also elaborate merit and demerit of existing works. 
Bansal et al. \cite{2020Lightweight} proposed a lightweight authentication and key agreement protocol based on physically uncloneable function (PUF) for V2G. Firstly, the challenge response and key agreement are applied to achieve mutual authentication and prevent the aggregator from being captured, respectively.
The EV authenticates with the aggregator via a pseudonym.
Then the EV and aggregator respectively select a random number to embed in public non-linear function.
Finally, the EV and the aggregator generate session keys respectively to ensure communication security.

\begin{table*}\small
	\caption{Authentication Protocols in V2G Charging}\label{RWV2G}
	\centering
	\begin{tabularx}{\textwidth}{cXX}
		\toprule
		Scheme & Primitives & Merit and Demerit\\
		\midrule
		A lightweight authentication \\and key agreement protocol \cite{2020Lightweight} &
		\begin{itemize}
			\item PUF
			\item Message authentication code (MAC)/HMAC
			\item Encryption/Decryption
			\item XOR operation
		\end{itemize} &
		\begin{itemize}
			\item Lightweight, privacy preserving, and secure.
			\item Only logical analysis. Smart Grid must be trusted so that it can not resist internal attacks. Too many times of encryption/decryption.
		\end{itemize}\\
		A mutual \\ authentication scheme\cite{2012Aggregated} &
		\begin{itemize}
			\item HMAC
			\item Symmetric encryption
			\item Bitwise logical operation
			\item Pseudo-random number generation
			\item Defined arithmetic functions
		\end{itemize} &
		\begin{itemize}
			\item Resist major attacks, such as impersonation attacks, replay attacks, and DoS attacks.
			\item Lack of comparison with existing works. There are many bitwise logic operations for each entity.
		\end{itemize}\\
		An anonymous \\ authentication scheme\cite{2015An} &
		\begin{itemize}
			\item HMAC
			\item Hash function
			\item Bilinear pairing
			\item Vector commitment
			\item Group signatures
		\end{itemize} &
		\begin{itemize}
			\item Dynamically revoke users, aggregation, and batch verification.
			\item The revocation list needs to be updated frequently. Generating certificates for each member consumes lots of overheads. AHO signature generated when revoking members has large overheads.
		\end{itemize}\\
		\bottomrule
	\end{tabularx}
\end{table*}

\begin{table*}\small
	\caption{Authentication Protocols in V2G Charging}\label{RWV2G2}
	\centering
	\begin{tabularx}{\textwidth}{cXX}
		\toprule
		Scheme & Primitives & Merit and Demerit\\
		\midrule		
		An authentication protocol\cite{2016Authentication} &
		\begin{itemize}
			\item Hash function
			\item Bilinear pairing
		\end{itemize} &
		\begin{itemize}
			\item Internal attacks resistance.
			\item Large overheads in the visiting V2G network.
		\end{itemize}\\
		A lightweight authentication\\ protocol  curve\cite{2020A} &
		\begin{itemize}
			\item Hash function
			\item Non-supersingular elliptic curve
			\item Two-party computation protocol
		\end{itemize} &
		\begin{itemize}
			\item Internal attacks resistance and lightweight.
			\item Unable to resist MITM attacks.
		\end{itemize}\\
		A lightweight authentication \\and key agreement\cite{9504586} &
		\begin{itemize}
			\item One-way hash functions
			\item XOR operation
			\item Signcryption
		\end{itemize} &
		\begin{itemize}
			\item Low computing overhead, suitable for resource-constrained environments.
			\item Unable to resist internal attacks and energy overheads are large.
		\end{itemize}\\
An enhanced mutually \\authenticated security protocol \cite{111111} &
		\begin{itemize}
			\item One-way hash functions
			\item XOR operation
		\end{itemize} &
		\begin{itemize}
			\item Ensure secure communication among EV, CS, and service providers.
			\item Formal proof is ignored.
		\end{itemize}\\
		\bottomrule
	\end{tabularx}
\end{table*}

Liu et al. \cite{2012Aggregated} proposed mutual authentication schemes for V2G under home mode and visiting mode, in which virtual battery vehicle (VBV) assists aggregators in validating EVs.
Under home mode, VBV is responsible for generating pseudonym and session identification.
Under visiting mode, VBV generates parameters that assist the authentication between EVs and aggregator.
In addition, the aggregator generates the parameter pseudo-status to achieve dynamic joining and leaving of the EV.

According to the V2G architecture of the Chinese market \cite{2015An}, provincial power service providers are regarded as central aggregators (CAGs), and municipal power service providers are regarded as local aggregators (LAGs).
The CAG allocates licenses to EV for authentication, and the subset covering framework is applied to revoke departing users.
In other words, the revoked recipients are divided into a group, while the unrevoked users are divided into different subsets.
The scheme applies a strongly unforgeable one-time signature to prevent attackers from associating the license with the real identity of EV.

Saxena et al. \cite{2016Authentication} proposed an authentication protocol for centralized charging and distributed charging in V2G, respectively.
For centralized charging, EV first hashes the pseudonym and timestamp, then sends the result and a selected random number to the aggregator. 
Subsequently, the aggregator sends the pseudonym of EV, the random number selected by EV, and other parameters to the certificate authority (CA). 
For distributed charging, in the home mode, the EV authenticates home CA through home aggregator.
In the multi-visiting mode, to prevent the attacker from using the home charging pile as a legitimate user, a two-factor authentication is utilized to realize malicious node tracking and anonymous authentication.

Su et al. \cite{2020A} proposed a lightweight authentication protocol based on non-supersingular elliptic curve, in which the third party and dispatching center generate the master key through two-party computation protocol to solve the issue of master key leakage.
Specifically, EV and CS firstly register at RSU.
Then, EV and CS embed their own selected random numbers in the parameters.
Finally, EV and CS negotiate to generate public parameters.
Under the random oracle model, the protocol is proven to be secure based on the elliptic curve discrete logarithm problem (ECDLP).
In addition, the comparison experiment also shows that the performance of the scheme is excellent, but the scheme can not resist MITM attacks.

Ahmed et al. \cite{9504586} proposed a lightweight authentication and key agreement protocol via signcryption and hash operation. 
The electrical service provider (ESP) embeds the random number, its own identity and identity of CS into the shared key for secure communication between ESP and CS. 
EV and CS complete mutual authentication with the assistance of ESP. 
Specifically, ESP authenticates the identities of CS and EV, by verifying that the key-hash response sent from CS and EV. 
Once the authentication is successful, ESP generates session keys for CS and EV to ensure secure energy trading between CS and EV.

To resist traceability attacks and impersonation attacks in V2G, Sureshkumar et al. \cite{111111} designed an authentication protocol based on elliptic curve cryptography (ECC) to achieve mutual authentication among EV, CS, and service providers. 
Additionally, the session key is generated to ensure secure communication among EV, CS, and service providers, whereas a formal proof is ignored.

\subsubsection{Authentication Protocols for V2V Charging}
Table \ref{RWV2V} reviews authentication protocols in V2V charging, and elaborates merit and demerit of existing works.
Roberts et al. \cite{2017An} proposed an authentication protocol for V2V charging based on mutual response.
To resist the MITM attack, the shared key is negotiated through Diffie-Hellman key exchange. 
Then the shared key and the random number are applied to finish V2V charging authentication. 
Finally, two Android phones are applied to simulate the charging demand side and the discharging side. 
A notebook is utilized to simulate the dedicated short range communication unit. 
Although the scheme can resist MITM attacks, it ignores security proof and location privacy protection. 
Additionally, the scheme can not resist other attacks, such as eavesdropping attacks.

Baza et al. \cite{9492014} proposed a privacy protection scheme for charging-station-to-vehicle (CS2V) and V2V charging based on blockchain. 
In V2V charging privacy protection scheme, prefix linkable anonymous authentication is applied to realize the anonymous authentication of EV.
To prevent double-spending and stolen coin spending attacks, zero knowledge proof is applied to verify currency ownership in batches on the premise of ensuring privacy.
The shared key is applied to resist spoofing attacks.
The transparency and unforgeability of blockchain are applied to prevent attackers from forging secret parameters.
However, with the increase of the number of EV, the public blockchain suffers from high gas consumption and data storage redundancy.

To prevent energy aggregators from leaking privacy by analyzing user data, Almuhaideb and Algothami \cite{222222} proposed a privacy protection authentication protocol based on elliptic curve qu-vanstone (ECQV) for V2V charging. 
The proposed scheme utilizes the credentials generated by ECQV to authenticate EV, so as to balance anonymity and traceability. Nonetheless, the communication overhead of this protocol is large.

\begin{table}[bp]\small
	\caption{Authentication Protocols in V2V Charging}\label{RWV2V}
	\centering
	
	\begin{tabularx}{\textwidth}{cXX}
		\toprule
		Scheme & Primitives & Merit and Demerit\\
		\midrule
		\cite{2017An} &
		\begin{itemize}
			\setlength{\itemsep}{0pt}			
			\setlength{\parsep}{0pt}
			\setlength{\parskip}{0pt}
			\item Key exchange
			\item AES encryption
			\item MD5 and SHA-256 hash functions
		\end{itemize} &
		\begin{itemize}
			\setlength{\itemsep}{0pt}			
			\setlength{\parsep}{0pt}
			\setlength{\parskip}{0pt}
			\item MITM attacks resistance.
			\item Lack of formal security proof. Location privacy leakage.
		\end{itemize}\\
		
		\cite{9492014} &
		\begin{itemize}
			\setlength{\itemsep}{0pt}			
			\setlength{\parsep}{0pt}
			\setlength{\parskip}{0pt}
			\item Prefix linkable anonymous authentication
			\item Blockchain
			\item Zero knowledge proof
		\end{itemize} &
		\begin{itemize}
			\setlength{\itemsep}{0pt}			
			\setlength{\parsep}{0pt}
			\setlength{\parskip}{0pt}
			\item Support privacy-preserving.
			\item Gas consumption increases as the number of EV increases.
		\end{itemize}\\

      \cite{222222} &
		\begin{itemize}
			\setlength{\itemsep}{0pt}			
			\setlength{\parsep}{0pt}
			\setlength{\parskip}{0pt}
			\item Elliptic curve qu-vanstone
		\end{itemize} &
		\begin{itemize}
			\setlength{\itemsep}{0pt}			
			\setlength{\parsep}{0pt}
			\setlength{\parskip}{0pt}
			\item Balance anonymity and traceability.
			\item Communication overhead is large.
		\end{itemize}\\
		\bottomrule
	\end{tabularx}
\end{table}

\subsubsection{Authentication Protocols for Dynamic Charging}
Table \ref{RWDC} reviews authentication protocols in dynamic charging, and elaborates merit and demerit of existing works. 
Ponnuru et al. \cite{9721573} proposed lightweight authentication and key agreement protocols, for dynamic charging systems based on fog computing architecture.
Specifically, RSU acts as an access point for dynamic charging.
EV and RSU finish mutual authentication via hash operation and exclusive OR (XOR) operation.
After passing the authentication, EV and CP within the range of RSU will finish mutual authentication.
This scheme proposes a handover authentication to improve the efficiency of EV switching between adjacent RSUs.
Although the computational overhead of this scheme is small, it is with a large communication overhead.

Badu et al. \cite{9552561} proposed a robust authentication protocol for dynamic charging based on cloud and fog architecture.
For mutual authentication, the EV and RSU embed the random number generated by the fog server into hash-based message authentication code (HMAC).
Then, RSU hashes the session key to generate the verification key, which is sent to EV to verify the validity of session key.
After mutual authentication between EV and RSU is successful, the RSU applies the session key to encrypt the random seed, which is sent to the EV and CPs.
However, the computing overhead of RSU is large.
Additionally, this protocol assumes that the cloud server is completely trustworthy, leading to a single point of collapse.

Li et al. \cite{7404052} applied pseudonyms to provide location privacy protection for EVs, and improved the efficiency of authentication through symmetric keys.
To balance the security and overhead of the authentication protocol, the session key is generated for CP during low traffic flow in advance, to prevent the slow speed of key generation when the traffic flow is dense.
Inspired by the roaming charging mode of smart phones, EV owners can subscribe to service providers according to their needs.
The authentication protocol is deployed in Raspberry Pi-2 for experimental simulation.
Results showed that the verification efficiency was significantly improved.

Under fog and cloud architecture, Roman et al. \cite{2020Authentication} proposed a charging request authentication scheme for dynamic charging.
EVs and RSUs realize mutual authentication through the fog server.
After the authentication is passed, RSU broadcasts a message to the CP indicating that the EV can be charged.
The identity anonymity of EV is protected by purchasing tickets from the charging service provider.
The scheme realizes secure payment and anonymous authentication, but the communication overhead will increase sharply with the increase of the number of CPs.

To improve the timeliness of authentication in dynamic charging, Pazos-Revilla et al. \cite{2017Secure} proposed an authentication scheme based on physical layer, in which EV uses a shared key to authenticate with RSU. 
EV and CP are authenticated by secret seeds. 
The scheme is simulated under various transmission power, and the transmission power is dynamically adjusted according to different weather. 
Finally, the optimal transmission power against eavesdropping is selected. 
This scheme realizes the secure and efficient authentication of the physical layer, but still lacks the formal security proof.

To ensure efficient dynamic charging and payment of authorized EV, Alshaeri et al. \cite{333333} applied stream cipher and blockchain to realize authentication between EV and CP within 0.051 milliseconds. Stream cipher is utilized to generation tokens, while blockchain is utilized to assist charging service provider in energy trade (i.e. bidding, charging, and billing). 
However, tokens used for authentication inevitably consume computational overhead.

\begin{table*}[!htbp]\small
	\caption{Authentication Protocols in Dynamic Charging}\label{RWDC}
	\centering
	
	\begin{tabularx}{\textwidth}{cXX}
		\toprule
		Scheme & Primitives & Merit and Demerit\\
		\midrule
		\cite{9721573} &
		\begin{itemize}
			\setlength{\itemsep}{0pt}			
			\setlength{\parsep}{0pt}
			\setlength{\parskip}{0pt}
			\item Hash functions
			\item Message Authentication Code
			\item XOR operation
		\end{itemize} &
		\begin{itemize}
			\setlength{\itemsep}{0pt}			
			\setlength{\parsep}{0pt}
			\setlength{\parskip}{0pt}
			\item Lightweight authentication protocol.
			\item High communication overheads.
		\end{itemize}\\
		
		\cite{9552561} &
		\begin{itemize}
			\setlength{\itemsep}{0pt}			
			\setlength{\parsep}{0pt}
			\setlength{\parskip}{0pt}
			\item ECC
			\item Hash function
			\item XOR operation
			\item HMAC
		\end{itemize} &
		\begin{itemize}
			\setlength{\itemsep}{0pt}			
			\setlength{\parsep}{0pt}
			\setlength{\parskip}{0pt}
			\item Total costs are low.
			\item RSUs have high computational overheads and charging server is threatened with single point of collapse.
		\end{itemize}\\
		
		\cite{7404052} &
		\begin{itemize}
			\setlength{\itemsep}{0pt}			
			\setlength{\parsep}{0pt}
			\setlength{\parskip}{0pt}
			\item AES encryption
			\item Digital signature
		\end{itemize} &
		\begin{itemize}
			\setlength{\itemsep}{0pt}			
			\setlength{\parsep}{0pt}
			\setlength{\parskip}{0pt}
			\item Support location privacy protection and key pre-distribution. Authentication efficiency has been greatly improved.
			\item Lack of formal security proof.
		\end{itemize}\\
		
		\cite{2020Authentication} &
		\begin{itemize}
			\setlength{\itemsep}{0pt}			
			\setlength{\parsep}{0pt}
			\setlength{\parskip}{0pt}
			\item Blind signature
			\item Bilinear pairing
			\item Hash functions
		\end{itemize} &
		\begin{itemize}
			\setlength{\itemsep}{0pt}			
			\setlength{\parsep}{0pt}
			\setlength{\parskip}{0pt}
			\item Support secure payment and location privacy protection.
			\item The number of messages increases linearly with the increase in the number of charging tablets.
		\end{itemize}\\
		
		\cite{2017Secure} &
		\begin{itemize}
			\setlength{\itemsep}{0pt}			
			\setlength{\parsep}{0pt}
			\setlength{\parskip}{0pt}
			\item Hash function
			\item XOR operation
			\item Secret sharing
		\end{itemize} &
		\begin{itemize}
			\setlength{\itemsep}{0pt}			
			\setlength{\parsep}{0pt}
			\setlength{\parskip}{0pt}
			\item Realize efficient authentication and privacy protection in the physical layer.
			\item Lack of formal security proof.
		\end{itemize}\\

        \cite{333333} &
		\begin{itemize}
			\setlength{\itemsep}{0pt}			
			\setlength{\parsep}{0pt}
			\setlength{\parskip}{0pt}
			\item Stream cipher
			\item Blockchain
		\end{itemize} &
		\begin{itemize}
			\setlength{\itemsep}{0pt}			
			\setlength{\parsep}{0pt}
			\setlength{\parskip}{0pt}
			\item Support efficient dynamic charging and payment of authorized EV.
			\item Tokens consume high computational overhead.
		\end{itemize}\\
		\bottomrule
	\end{tabularx}
\end{table*}

\subsection{Encryption}
\subsubsection{Homomorphic Encryption}
Table \ref{RWhe} compares related works in homomorphic encryption. 
Here, Chen et al. \cite{9523794} introduced EC and full homomorphic encryption, to ensure the confidentiality and privacy of traffic data transmitted by edge nodes.
In addition, distributed edge nodes apply ciphertext to train model parameters, protecting the privacy of model parameters.
Finally, parameters of model are stored on the blockchain for updating the model parameters robustly and reliably.
\begin{table*}[!htbp]\small
	\caption{Related Works in Homomorphic Encryption}\label{RWhe}
	\centering
	
	\begin{tabularx}{\textwidth}{cXX}
		\toprule
		Scheme & Primitives & Merit and demerit\\
		\midrule
		\cite{9523794} &
		\begin{itemize}
			\setlength{\itemsep}{0pt}			
			\setlength{\parsep}{0pt}
			\setlength{\parskip}{0pt}
			\item Full homomorphic encryption
			\item Blockchain
		\end{itemize} &
		\begin{itemize}
			\setlength{\itemsep}{0pt}			
			\setlength{\parsep}{0pt}
			\setlength{\parskip}{0pt}
			\item The privacy of model parameters is protected, ensuring a trusted, asynchronous, and decentralized model update.
			\item Low computational efficiency
		\end{itemize}\\
		
		\cite{9344812} &
		\begin{itemize}
			\setlength{\itemsep}{0pt}			
			\setlength{\parsep}{0pt}
			\setlength{\parskip}{0pt}
			\item Lattice-based full homomorphic encryption
		\end{itemize} &
		\begin{itemize}
			\setlength{\itemsep}{0pt}			
			\setlength{\parsep}{0pt}
			\setlength{\parskip}{0pt}
			\item Protect a theoretical basis for post quantum cryptography.
			\item Secure access to private data sacrifices overheads because of the introduction of a secret bootstrapping key.
		\end{itemize}\\
		
		\cite{2017A} &
		\begin{itemize}
			\setlength{\itemsep}{0pt}			
			\setlength{\parsep}{0pt}
			\setlength{\parskip}{0pt}
			\item Paillier cryptosystem
			\item Chinese Remainder Theorem
			\item Proxy re-encryption
			\item Hash functions
		\end{itemize} &
		\begin{itemize}
			\setlength{\itemsep}{0pt}			
			\setlength{\parsep}{0pt}
			\setlength{\parskip}{0pt}
			\item Support privacy-preserving sensory data sharing.
			\item Unable to resist MITM attacks and sybil attacks. Fewer experimental comparisons and lack of formal security proof.
		\end{itemize}\\
		
		\cite{2016IP2DM} &
		\begin{itemize}
			\setlength{\itemsep}{0pt}			
			\setlength{\parsep}{0pt}
			\setlength{\parskip}{0pt}
			\item Access control
			\item Homomorphic encryption
			\item Onion-level encryption
		\end{itemize} &
		\begin{itemize}
			\setlength{\itemsep}{0pt}			
			\setlength{\parsep}{0pt}
			\setlength{\parskip}{0pt}
			\item Support layered privacy protection.
			\item Operation time per transaction is large.
		\end{itemize}\\
		
		\cite{2020The} &
		\begin{itemize}
			\setlength{\itemsep}{0pt}			
			\setlength{\parsep}{0pt}
			\setlength{\parskip}{0pt}
			\item Full homomorphic encryption
			\item Blockchain
		\end{itemize} &
		\begin{itemize}
			\setlength{\itemsep}{0pt}			
			\setlength{\parsep}{0pt}
			\setlength{\parskip}{0pt}
			\item Protect data confidentiality.
			\item The efficiency of decryption is low.
		\end{itemize}\\

        \cite{444444} &
		\begin{itemize}
			\setlength{\itemsep}{0pt}			
			\setlength{\parsep}{0pt}
			\setlength{\parskip}{0pt}
			\item Homomorphic encryption
		\end{itemize} &
		\begin{itemize}
			\setlength{\itemsep}{0pt}			
			\setlength{\parsep}{0pt}
			\setlength{\parskip}{0pt}
			\item Protect the privacy of vehicle data.
			\item The data pre-processing and encryption are time-consuming.
		\end{itemize}\\
		\bottomrule
	\end{tabularx}
\end{table*}

Karim and Rawat \cite{9344812} proposed a privacy risk assessment model to prevent attackers from tracking vehicles through toll transponders.
The model comprehensively considers privacy impact, breach impact, and shareability impact to classify security threats.
According to the threat level, the appropriate encryption level is utilized to protect the driver's privacy.
Finally, the scheme proposed a lattice-based full homomorphic encryption to verify the correctness of the risk assessment model, and this provides a theoretical basis for post quantum cryptography.

To prevent collusion between the trusted center and RSU, as well as privacy disclosure caused by vehicle crowdsensing, Kong et al. \cite{2017A} proposed a privacy protection scheme without the participation of the trusted center. 
RSUs encrypt data via the improved Paillier cryptosystem, then the ciphertext is transformed into re-encrypted ciphertext for retrieval.

According to the different security requirements of V2G management, Han and Xiao \cite{2016IP2DM} integrated access control, data aggregation, onion-level encryption and other technologies to protect users' privacy.
In onion encryption, the homomorphic encryption is applied to safely calculate the data stored in the database and aggregate power data. 
Compared with data aggregation without homomorphic encryption, the scheme achieves great advantages in throughput and time cost.

To protect the confidentiality of data interaction between cloud center and vehicle manufacturer, Cui et al. \cite{2020The} applied homomorphic encryption to process data.
Moreover, the key switch is applied to alleviate the sharp increase of ciphertext dimension, and reduce the noise introduced by ciphertext calculation.

Boudguiga et al. \cite{444444} protected the privacy of vehicle data via homomorphic encryption. Due to delays in the remote cloud, vehicles offload computing tasks to the edge server, then server classifies driving behavior via a three-layer neural network. 
Nevertheless, the data pre-processing and encryption are time-consuming.

\subsubsection{Broadcast Encryption}

\begin{table*}[!htbp]\small
	\caption{Related Works in Broadcast Encryption}\label{RWBE}
	\centering
	
	\begin{tabularx}{\textwidth}{cXX}
		\toprule
		Scheme & Primitives & Merit and Demerit\\
		\midrule
		\cite{9541084} &
		\begin{itemize}
			\setlength{\itemsep}{0pt}			
			\setlength{\parsep}{0pt}
			\setlength{\parskip}{0pt}
			\item Identity-based broadcast encryption
			\item Proxy re-encryption
			\item Hash function
			\item Bilinear pairing
		\end{itemize} &
		\begin{itemize}
			\setlength{\itemsep}{0pt}			
			\setlength{\parsep}{0pt}
			\setlength{\parskip}{0pt}
			\item Total costs is low
			\item Overheads of proxy server are large. Lack of formal security proof.
		\end{itemize}\\	
		\cite{2020VANET} &
		\begin{itemize}
			\setlength{\itemsep}{0pt}			
			\setlength{\parsep}{0pt}
			\setlength{\parskip}{0pt}
			\item Group broadcast encryption
		\end{itemize} &
		\begin{itemize}
			\setlength{\itemsep}{0pt}			
			\setlength{\parsep}{0pt}
			\setlength{\parskip}{0pt}
			\item Ensure secure communication among group members.
			\item Details on bridging groups are lacking.
		\end{itemize}\\		
		\cite{2018Anonymous} &
		\begin{itemize}
			\setlength{\itemsep}{0pt}			
			\setlength{\parsep}{0pt}
			\setlength{\parskip}{0pt}
			\item Bilinear pairing
			\item Certificate-based broadcast encryption
		\end{itemize} &
		\begin{itemize}
			\setlength{\itemsep}{0pt}			
			\setlength{\parsep}{0pt}
			\setlength{\parskip}{0pt}
			\item Solve the problem of certificate management and decryption cost is constant.
			\item The ciphertext size increases linearly with the increase of the number of receivers.
		\end{itemize}\\	
		\cite{2020Anonymous} &
		\begin{itemize}
			\setlength{\itemsep}{0pt}			
			\setlength{\parsep}{0pt}
			\setlength{\parskip}{0pt}
			\item Access control
			\item Certificate-based broadcast encryption
			\item Bilinear pairing
		\end{itemize} &
		\begin{itemize}
			\setlength{\itemsep}{0pt}			
			\setlength{\parsep}{0pt}
			\setlength{\parskip}{0pt}
			\item Solve key escrow issues. The system parameter size, key length, and decryption cost are constant.
			\item The ciphertext size increases linearly with the increase of the number of receivers.
		\end{itemize}\\		
		\cite{2019Revocable} &
		\begin{itemize}
			\setlength{\itemsep}{0pt}			
			\setlength{\parsep}{0pt}
			\setlength{\parskip}{0pt}
			\item Identity-based broadcast encryption
			\item Proxy re-encryption
			\item Bilinear pairing
			\item Hash function
		\end{itemize} &
		\begin{itemize}
			\setlength{\itemsep}{0pt}			
			\setlength{\parsep}{0pt}
			\setlength{\parskip}{0pt}
			\item Realize the flexible revocation of visitors.
			\item Storing and updating delegatee lists and revocation list requires significant overheads.
		\end{itemize}\\	
		\cite{39} &
		\begin{itemize}
			\setlength{\itemsep}{0pt}			
			\setlength{\parsep}{0pt}
			\setlength{\parskip}{0pt}
			\item Hash function
			\item Bilinear pairing
			\item Direct revocation
			\item Lagrange interpolation
		\end{itemize} &
		\begin{itemize}
			\setlength{\itemsep}{0pt}			
			\setlength{\parsep}{0pt}
			\setlength{\parskip}{0pt}
			\item Dynamically adjust data access.
			\item Revocation list requires significant overhead.
		\end{itemize}\\

        \cite{555555} &
		\begin{itemize}
			\setlength{\itemsep}{0pt}			
			\setlength{\parsep}{0pt}
			\setlength{\parskip}{0pt}
			\item Identity-based signcryption
		\end{itemize} &
		\begin{itemize}
			\setlength{\itemsep}{0pt}			
			\setlength{\parsep}{0pt}
			\setlength{\parskip}{0pt}
			\item Realize one-to-many data sharing.
			\item The issue of key escrow is introduced.
		\end{itemize}\\		
		\bottomrule
	\end{tabularx}
\end{table*}

Table \ref{RWBE} compares related works in broadcast encryption.
Zhong et al. \cite{9541084} proposed identity-based broadcast encryption to solve data redundancy. 
Vehicles in the group initially register with the trusted authority (TA), then those have passed identity authentication will receive the ciphertext encrypted by TA.
To reduce the computational burden of TA, the proxy server converts the ciphertext into re-encrypted ciphertext, to realize secure and efficient one-to-many communication.
Bunese et al. \cite{2020VANET} further proposed a group broadcast encryption for vehicle communication, by dividing network data into the plaintext, group messages, and secure group messages.

In view of certificate management issues in the existing broadcast encryption based on public key cryptosystem, Li et al. \cite{2018Anonymous} proposed a broadcast encryption based on anonymous certificate, with suitability for large-scale receivers with limited computing power. 
Under the random oracle, the scheme is proven to meet the adaptive CCA-secure with the computational bilinear Diffie Hellman assumption. 
More importantly, the decryption time was constant. However, the communication overhead of ciphertext increases linearly with the number of receivers.

To solve the key leakage caused by the key escrow in identity-based cryptosystem and protect the anonymity of authorized users, Chen et al. \cite{2020Anonymous} proposed a broadcast encryption with identity anonymity.
Under the standard model, the scheme is proven to meet the adaptive CCA2-secure with q-augmented bilinear Diffie-Hellman exponent assumption and 1-bilinear Diffie-Hellman exponent inversion assumption. 
More importantly, the communication overhead of system parameters and keys, and decryption time were constant. 
However, the length of ciphertext increases linearly with the number of receivers.

Ge et al. \cite{2019Revocable} proposed an identity-based broadcast encryption that supports key revocation to realize flexible data outsourcing encryption and revocation. Users apply a random polynomial to randomize their private keys, to resist the collusion between agents and attackers. 
Delegatee list (denoted by S) and revocation list (denoted by R) are applied to realize data access control. 
Subtracting R from S is a new delegatee list. Under the random oracle, based on (f, g, F) - general decision Diffie-Hellman exponet (GDDHE) assumption \cite{2016Recipient}, the scheme is proven to CPA-secure. However, storing and updating lists still require significant overheads.

Aiming at the low efficiency of user authority revocation in smart grid, Niu et al. \cite{39} proposed an identity-based broadcast encryption scheme for managing user authority. 
This scheme is proven to be semantically secure under the random oracle, and realizes secure data transmission of smart grid. However, the revocation list inevitably leads to large overheads.

Zhao et al. \cite{555555} proposed a broadcast signcryption scheme for vehicular platoon. The platoon leader generates a fixed-length instruction based on identity-based signcryption to realize one-to-many data sharing. 
The communication overhead of ciphertext is lightweight because the length of ciphertext does not increase with the number of platoon members. 
However, the issue of key escrow is introduced.

\subsubsection{Attribute-Based Encryption}
\begin{table*}[!htbp]\small
	\caption{Related Works in Attribute-based Encryption}\label{RWAB111}
	\centering
	
	\begin{tabularx}{\textwidth}{cXX}
		\toprule
		Scheme & Primitives & Merit and demerit\\
		\midrule
		\cite{2020Attribute} &
		\begin{itemize}
			\setlength{\itemsep}{0pt}			
			\setlength{\parsep}{0pt}
			\setlength{\parskip}{0pt}
			\item Tree access structure
			\item Linear secret-sharing
			\item Bilinear pairing
		\end{itemize} &
		\begin{itemize}
			\setlength{\itemsep}{0pt}			
			\setlength{\parsep}{0pt}
			\setlength{\parskip}{0pt}
			\item Secure parallel outsourced decryption
			\item High computational complexity.
		\end{itemize}\\
		
		\cite{2019SCTSC} &
		\begin{itemize}
			\setlength{\itemsep}{0pt}			
			\setlength{\parsep}{0pt}
			\setlength{\parskip}{0pt}
			\item Blockchain
			\item Hash functions
		\end{itemize} &
		\begin{itemize}
			\setlength{\itemsep}{0pt}			
			\setlength{\parsep}{0pt}
			\setlength{\parskip}{0pt}
			\item Attributes are divided into static attributes and dynamic attributes to realize fine-grained access.
			\item Along with the increase of attributes, the encryption time is longer.
		\end{itemize}\\
		
		\cite{9523580} &
		\begin{itemize}
			\setlength{\itemsep}{0pt}			
			\setlength{\parsep}{0pt}
			\setlength{\parskip}{0pt}
			\item Attribute-based encryption based on ciphertext policy
			\item Linear secret-sharing
			\item Access structure
		\end{itemize} &
		\begin{itemize}
			\setlength{\itemsep}{0pt}			
			\setlength{\parsep}{0pt}
			\setlength{\parskip}{0pt}
			\item Support fine-grained access and dynamic revocation of vehicles.
			\item Computing overheads of edge nodes are not evaluated.
		\end{itemize}\\
		
		\cite{2021Attribute} &
		\begin{itemize}
			\setlength{\itemsep}{0pt}			
			\setlength{\parsep}{0pt}
			\setlength{\parskip}{0pt}
			\item Access tree
			\item Blockchain
			\item Bilinear pairing
		\end{itemize} &
		\begin{itemize}
			\setlength{\itemsep}{0pt}			
			\setlength{\parsep}{0pt}
			\setlength{\parskip}{0pt}
			\item Solve access control of the announcement message on the VANETs.
			\item Ignore the efficiency of data verification and retrieval in the blockchain, as well as the gas consumption of smart contract.
		\end{itemize}\\
		\cite{EltayiebEHL19} &
		\begin{itemize}
			\setlength{\itemsep}{0pt}			
			\setlength{\parsep}{0pt}
			\setlength{\parskip}{0pt}
			\item Hash function
			\item Bilinear pairing
			\item Bilinear pairing
		\end{itemize} &
		\begin{itemize}
			\setlength{\itemsep}{0pt}			
			\setlength{\parsep}{0pt}
			\setlength{\parskip}{0pt}
			\item Realize secure data retrieval.
			\item Ciphertext size is large.
		\end{itemize}\\	
		\cite{45} &
		\begin{itemize}
			\setlength{\itemsep}{0pt}			
			\setlength{\parsep}{0pt}
			\setlength{\parskip}{0pt}
			\item Hash function
			\item XOR operation
		\end{itemize} &
		\begin{itemize}
			\setlength{\itemsep}{0pt}			
			\setlength{\parsep}{0pt}
			\setlength{\parskip}{0pt}
			\item Support multi-attribute-based secure communication.
			\item Lack of formal security proof and communication cost is large.
		\end{itemize}\\

	   \cite{666666} &
		\begin{itemize}
			\setlength{\itemsep}{0pt}			
			\setlength{\parsep}{0pt}
			\setlength{\parskip}{0pt}
			\item Ciphertext polity attribute-based encryption
		\end{itemize} &
		\begin{itemize}
			\setlength{\itemsep}{0pt}			
			\setlength{\parsep}{0pt}
			\setlength{\parskip}{0pt}
			\item Realize the secure access control.
			\item The encryption efficiency is low.
		\end{itemize}\\
		\bottomrule
	\end{tabularx}
\end{table*}

Table \ref{RWAB111} compares related works in attribute-based encryption. 
Existing attribute-based encryption uses time-consuming bilinear mapping, resulting in slow computing speed.
Feng et al. \cite{2020Attribute} proposed an attribute-based encryption scheme, which applies tree access structure to support parallel outsourcing decryption based on Spark and MapReduce framework. 
To improve the universality of scheme, the linear secret sharing is converted into an access tree structure.

To efficiently and dynamically regulate traffic lights, Cheng et al. \cite{2019SCTSC} proposed attribute-based encryption supporting fine-grained access to balance privacy and availability.
Here, attributes are divided into static attributes (such as inherent attributes) and dynamic attributes (such as modifiable attributes).

Zhao et al. \cite{9523580} proposed an attribute-based encryption scheme based on ciphertext policy, to realize efficient revocation of vehicles for vehicular platoon. 
Part of the decryption task is outsourced to edge nodes, to reduce the computing overhead of vehicles. 
Finally, jamming attacks, velocity fluctuation and distance fluctuation are applied to evaluate the effectiveness of scheme.

To support fine-grained access according to users' interests and attributes, Ma et al. \cite{2021Attribute} proposed an attribute-based encryption for VANETs.
To prevent single point of collapse caused by centralization, blockchain is applied to store shared traffic data. Nonetheless, this scheme ignores the efficiency of data verification and retrieval in the blockchain, as well as the gas consumption of smart contract.

To prevent privacy disclosure when data is outsourced to the cloud platform, Eltayieb et al. \cite{EltayiebEHL19} proposed an attribute-based searchable encryption scheme for smart grid. 
The scheme supports offline and online encryptions, in which offline encryption refers to the pre-processing without plaintext and attribute sets. 
Online encryption refers to the computation required for generating ciphertext when plaintext and attribute sets are known. 
The cloud center searches the data according to the trapdoor submitted by the user, to realize the secure access and privacy protection of the data. Nonetheless, the length of ciphertext does not have obvious advantage.

Chaudhary et al. \cite{45} proposed a secure multi-attribute communication scheme with attributed-based encryption. 
The cuckoo filter is applied to improve data forwarding efficiency, despite that the communication cost is still large. 
However, a formal security proof is ignored in this work.

Anish T. P et al. \cite{666666} proposed a weighted ciphertext polity attribute-based encryption, wherein blockchain is utilized to distributed verify, store and share data. 
Moreover, a hierarchical access control polity is provided for multi-role entities. 
The proposed scheme further realizes the secure access control, providing reliable data for forensics. However, the encryption efficiency is low.

\section{Network Management Security}\label{networkman}
Studies in Section \ref{datasecurity} only ensure the confidentiality of data, and the legality verification of network entity, with only defending against external attacks such as illegal access, traffic sniffing, and traffic analysis.
However, these studies cannot verify data from internal attackers with legitimate credentials, since these studies do not focus on the data and behavior of legal network entities.
Therefore, network management security (including trust management, misbehavior detection, IDS, and firewall) ensures the reliability, and the integrity and correctness of data exchanged within network system.

\subsection{Trust Management}\label{trust_manage}
\subsubsection{Trust Management for VANETs}
\textbf{Trust Mechanism Model:}
Due to limited capacity of vehicles, typical technologies requiring much higher computation resource such as intrusion detection, password encryption, and decryption technology are not applicable.
Alternatively, the purpose of trust mechanism is to evaluate the trust level of vehicles, based on their interaction history.
Using this as evaluation guidance, it is able to abandon malicious vehicles, and encourage trustworthy vehicles for data interaction.
Depending on the scope of trust model, the following four categories are introduced:

\textbf{Entity-based trust model:}
Literature based on entity-based trust model mainly evaluates the credibility of vehicles.
Here, the direct trust and indirect recommendation trust are applied jointly, to detect untrusted or malicious vehicles.
Tan et al. \cite{tan2015trust} applied graph theory to evaluate the routing trustworthyness of vehicles based on packet transmission rate and average delay.
Xiao et al. \cite{xiao2019bayestrust} proposed a trust model based on historical interaction between vehicles.
Such history is further used to construct a relatively stable trust link graph.
In addition, authors refer to PageRank in terms of calculating the trust value from global level, so as to unlock the cooperation potential of vehicles in trust management.

\textbf{Data-based trust model:} The data-based trust model aims to evaluate the reliability of data level.
Here, this trust model requires to collect data from a variety of sources, including vehicles themselves, their neighbors and RSUs.
Huang et al. \cite{huang2014social} proposed a novel voting mechanism based on the distance relationship between the vehicle and reported event.
Here, the vehicle decides whether to believe the received event data depending on voting result, where a closer distance to event implies a higher voting weight.
Besides, the signal strength and vehicle geographic location (e.g., GPS) are applied in literature \cite{rawat2015trust}.
This work further combines Bayesian estimation algorithm together with geographical determination to identify malicious data.

\textbf{Combined trust model:} The combined trust model by default, takes advantages of two types of trust models.
It not only evaluates the trust degree of vehicles, but also calculates the reliability of data.
Inherently, the motivation of combined trust model is that, the trust of vehicles affects the reliability of data due to the impact of interaction behavior, while the trust of data, in turn, reflects the reliability of vehicles due to the forwarding path that a data will be traversed.
The attack-resistant trust (ART) management \cite{li2015art} estimates the reliability of vehicles and data in a hybrid manner.
Firstly, ART applies the Dempster–Shafer evidence theory to evaluate the reliability of data.
Then, ART applies a collaborative filtering algorithm to calculate the reliability of vehicles, based on functional trust factor and recommendation trust factor.
Despite of high accuracy in terms of malicious vehicles detection, the performance of ART is influenced when there are insufficient number of vehicles participating the trust computing.
In addition, Farhan et al. \cite{ahmad2020marine} proposed a trust model to resist dedicated MITM attacks.
Firstly, the trust value of sender is calculated by a multi-dimensional entity-centered trust evaluation.
Once the receiver has verified the credibility of sender, sequentially the credibility of data is thus evaluated. 
Table \ref{trustman} lists related works in trust management.

\subsubsection{Trust Management in Smart Grid}
Fadul et al. \cite{Fadul2014ATrust} proposed a robust and configurable trust management toolkit, to protect smart grid from cyber attacks.
This trust management toolkit assigns different trust values to all smart grid devices (those to be protected in grid), where a low trust value indicates a higher security risk, and vice versa.
The toolkit combines reputation-based trust with network-flow algorithms, to identify and mitigate faulty smart grid protection nodes.
In addition to possible attacks, the toolkit also assigns a low trust value to faulty devices, indicating a high-risk failure that cannot be detected.

Gong et al. \cite{gong2018remote} proposed a sensing layer remote attestation mechanism to perceive the real-time trust of sensor nodes, and implemented trusted access control between sensor nodes.
Based on the formal description of sensor nodes, this mechanism proposes a real-time trust measure for sensor nodes.
Through the real-time tracking of node trust, this mechanism achieves the trusted communication between the terminal and control server node in the network environment. 
\begin{table*}[!htbp]\small
	\caption{Related Works in Trust Management}\label{trustman}
	\centering
	
	\begin{tabularx}{\textwidth}{cXX}
		\toprule
		Scheme & Primitives & Merit and Demerit\\
		\midrule
		\cite{tan2015trust} &
		\begin{itemize}
			\setlength{\itemsep}{0pt}			
			\setlength{\parsep}{0pt}
			\setlength{\parskip}{0pt}
			\item Fuzzy logic
                \item Game theory
		\end{itemize} &
		\begin{itemize}
			\setlength{\itemsep}{0pt}			
			\setlength{\parsep}{0pt}
			\setlength{\parskip}{0pt}
			\item Formulate imprecise empirical knowledge used for trust management
                \item Lack of adaptation ability
		\end{itemize}\\

		\cite{xiao2019bayestrust} &
		\begin{itemize}
			\setlength{\itemsep}{0pt}			
			\setlength{\parsep}{0pt}
			\setlength{\parskip}{0pt}
			\item PageRank
                \item Beyasian theory
		\end{itemize} &
		\begin{itemize}
			\setlength{\itemsep}{0pt}			
			\setlength{\parsep}{0pt}
			\setlength{\parskip}{0pt}
			\item Socially-aware trust model to reflect the interaction between vehicles
			\item Concern on randomness in trust management
		\end{itemize}\\

		\cite{huang2014social} &
		\begin{itemize}
			\setlength{\itemsep}{0pt}			
			\setlength{\parsep}{0pt}
			\setlength{\parskip}{0pt}
			\item Voting weight related to distance
		\end{itemize} &
		\begin{itemize}
			\setlength{\itemsep}{0pt}			
			\setlength{\parsep}{0pt}
			\setlength{\parskip}{0pt}
			\item Overcome information cascading and oversampling problem
                \item Lack of study on information delay for voting decision
		\end{itemize}\\

		\cite{rawat2015trust} &
		\begin{itemize}
			\setlength{\itemsep}{0pt}			
			\setlength{\parsep}{0pt}
			\setlength{\parskip}{0pt}
			\item Voting weight related to distance
		\end{itemize} &
		\begin{itemize}
			\setlength{\itemsep}{0pt}			
			\setlength{\parsep}{0pt}
			\setlength{\parskip}{0pt}
			\item Overcome information cascading and oversampling problem   
                \item Lack of study on information delay for voting decision
		\end{itemize}\\

		\cite{li2015art} &
		\begin{itemize}
			\setlength{\itemsep}{0pt}			
			\setlength{\parsep}{0pt}
			\setlength{\parskip}{0pt}
			\item Probabilistic estimation for vehicle level
			\item Deterministic estimation for data level
		\end{itemize} &
		\begin{itemize}
			\setlength{\itemsep}{0pt}			
			\setlength{\parsep}{0pt}
			\setlength{\parskip}{0pt}
			\item Combination of probabilistic and deterministic approach
			\item Lack of timely decision making response
		\end{itemize}\\	

            \cite{ahmad2020marine} &
		\begin{itemize}
			\setlength{\itemsep}{0pt}			
			\setlength{\parsep}{0pt}
			\setlength{\parskip}{0pt}
			\item Direct and indirect trust management
               \item Multi-dimensional plausibility checks
		\end{itemize} &
		\begin{itemize}
			\setlength{\itemsep}{0pt}			
			\setlength{\parsep}{0pt}
			\setlength{\parskip}{0pt}
			\item Identify MITM attacks as well as revokes their credentials
               \item Lack of investigation on social network impact
		\end{itemize}\\

            \cite{Fadul2014ATrust} &
		\begin{itemize}
			\setlength{\itemsep}{0pt}			
			\setlength{\parsep}{0pt}
			\setlength{\parskip}{0pt}
			\item Reputation-based trust
                \item Network-flow algorithms
		\end{itemize} &
		\begin{itemize}
			\setlength{\itemsep}{0pt}			
			\setlength{\parsep}{0pt}
			\setlength{\parskip}{0pt}
			\item Trust-management toolkit
               \item Automation level is the concern
		\end{itemize}\\

          \cite{gong2018remote} &
		\begin{itemize}
			\setlength{\itemsep}{0pt}			
			\setlength{\parsep}{0pt}
			\setlength{\parskip}{0pt}
			\item Signature
                \item Verification
		\end{itemize} &
		\begin{itemize}
			\setlength{\itemsep}{0pt}			
			\setlength{\parsep}{0pt}
			\setlength{\parskip}{0pt}
			\item Encapsulating the properties and trust value for tracing
               \item Low computation overhead
		\end{itemize}\\
		\bottomrule
	\end{tabularx}
\end{table*}
\subsection{Misbehavior Detection}\label{mis_detection}
\subsubsection{Detection Based on Message Content}
\begin{table*}[!htbp]\small
	\caption{Related Works in Misbehavior Detection}\label{misbehavior}
	\centering
	
	\begin{tabularx}{\textwidth}{cXX}
		\toprule
		Scheme & Primitives & Merit and Demerit\\
		\midrule
		\cite{jaeger2012novel} &
		\begin{itemize}
			\setlength{\itemsep}{0pt}			
			\setlength{\parsep}{0pt}
			\setlength{\parskip}{0pt}
			\item Location prediction
                \item Kalman filter
		\end{itemize} &
		\begin{itemize}
			\setlength{\itemsep}{0pt}			
			\setlength{\parsep}{0pt}
			\setlength{\parskip}{0pt}
                \item Accurately predict the movement of vehicles in the presence of errors.
			\item Violate the privacy of vehicle.
		\end{itemize}\\

		\cite{yang2019deqos} &
		\begin{itemize}
			\setlength{\itemsep}{0pt}			
			\setlength{\parsep}{0pt}
			\setlength{\parskip}{0pt}
			\item Distance-bounding
		\end{itemize} &
		\begin{itemize}
			\setlength{\itemsep}{0pt}			
			\setlength{\parsep}{0pt}
			\setlength{\parskip}{0pt}
                \item Accurately verify the location range of vehicle.
			\item High requirements for time delay.
		\end{itemize}\\

            \cite{alladi2021deepadv} &
		\begin{itemize}
			\setlength{\itemsep}{0pt}			
			\setlength{\parsep}{0pt}
			\setlength{\parskip}{0pt}
			\item Deep neural networks
		\end{itemize} &
		\begin{itemize}
			\setlength{\itemsep}{0pt}			
			\setlength{\parsep}{0pt}
			\setlength{\parskip}{0pt}
               \item DNNs are deployed on RSUs, avoiding the problem of limited vehicle resources.
			\item High RSUs requirements for the large data of VANETs.
		\end{itemize}\\

            \cite{gu2022cluster} &
		\begin{itemize}
			\setlength{\itemsep}{0pt}			
			\setlength{\parsep}{0pt}
			\setlength{\parskip}{0pt}
			\item Cluster
		\end{itemize} &
		\begin{itemize}
			\setlength{\itemsep}{0pt}			
			\setlength{\parsep}{0pt}
			\setlength{\parskip}{0pt}
               \item Consider the downstream data transmission procedure.
			\item Due to the high mobility of vehicles, frequent clustering is required.
		\end{itemize}\\
		
		\cite{dias2015cooperative} &
		\begin{itemize}
			\setlength{\itemsep}{0pt}			
			\setlength{\parsep}{0pt}
			\setlength{\parskip}{0pt}
			\item Watchdog
		\end{itemize} &
		\begin{itemize}
			\setlength{\itemsep}{0pt}			
			\setlength{\parsep}{0pt}
			\setlength{\parskip}{0pt}
                \item Evaluating the trust of neighbor vehicles based on their forwarding behavior.
			\item It does not consider the situation when the neighbor vehicle is malicious or the trust of neighbors is false.
		\end{itemize}\\

		\cite{li2012lie} &
		\begin{itemize}
			\setlength{\itemsep}{0pt}			
			\setlength{\parsep}{0pt}
			\setlength{\parskip}{0pt}
			\item Rate limit
		\end{itemize} &
		\begin{itemize}
			\setlength{\itemsep}{0pt}			
			\setlength{\parsep}{0pt}
			\setlength{\parskip}{0pt}
                \item Effectively alleviate flooding attacks.
			\item Can not alleviate DDoS.
		\end{itemize}\\

		\cite{yao2018multi} &
		\begin{itemize}
			\setlength{\itemsep}{0pt}			
			\setlength{\parsep}{0pt}
			\setlength{\parskip}{0pt}
			\item  Received signal strength indication
		\end{itemize} &
		\begin{itemize}
			\setlength{\itemsep}{0pt}			
			\setlength{\parsep}{0pt}
			\setlength{\parskip}{0pt}
			\item Consider the physical characteristics of vehicles that can not be hidden.
			\item Characteristics of RSSI may be exploited by attackers.
		\end{itemize}\\	
  
		\bottomrule
	\end{tabularx}
\end{table*}
Table \ref{misbehavior} compares related works in misbehavior detection mechanisms. Jaeger et al. \cite{jaeger2012novel} tracked the trajectory of vehicle by analyzing the sequence of cooperative awareness messages (CAMs).
By using the Kalman filter to track the vehicle, this work can predict the following position range of vehicle, thereby verifying the location contained within CAMs.
However, the Kalman filter can be interfered by multiple fake CAMs, thus the fake location contained in CAMs can fall within the prediction range.

Yang et al. \cite{yang2019deqos} proposed a scheme to verify the range of vehicles location based on the distance-bounding protocols.
Here, the approximate distance between the verifier and prover could be determined, by the round-trip time of cryptographic challenge-response pairs and the signal propagation speed.
If the sum of response error and timeout error is greater than the predefined threshold, the prover is considered to be out of the predefined distance range.

Alladi et al \cite{alladi2021deepadv} proposed an anomaly detection framework for VANETs based on deep neural networks (DNNs) with  applying a sequence reconstruction and thresholding algorithm.
The sequence reconstruction reconstructs the sequence that the messages extracted from the network (including faults, attacks and genuine messages) are converted into.
The thresholding algorithm calculates a potential threshold based on the error between the original sequence and reconstructed sequence.
Finally, this work classifies the sequence as anomalous or genuine based on the calculated potential threshold.
It is worth mentioning that their work deploys DNNs on RSUs to alleviate the problem of limited vehicle resources.

Gu et al \cite{gu2022cluster} proposed a cluster-based malicious vehicle detection mechanism for false downstream data in fog computing-based VANETs.
This work supervises the downstream data forwarded by cluster-head vehicles, and detects malicious cluster-head vehicles.
It first clusters vehicles based on their distance and link reliability, and selects cluster-head vehicles and edge monitoring vehicles.
Then the fog server is applied to detect suspicious data and record the suspicious cluster-head vehicles reported by edge monitoring vehicles in each cluster.
Finally, paired edge monitoring vehicles exchange and verify the data received from corresponding cluster heads to identify malicious cluster-head vehicles.

\subsubsection{Detection Based on Network Behavior}
In addition to the content of data received, malicious vehicles can also be detected based on the behavior of network participant, including forwarding, tampering, discarding, and selective discarding.
The advantage of mechanisms based on network behavior is completely independent of the data content.

Dias et al. \cite{dias2015cooperative} assigned an initial reputation score to qualify vehicle.
When the vehicle establishes contact with other vehicles, it will also receive the neighbor's reputation scores for other vehicles from the neighbor.
Then the reputation score of vehicle is updated based on its historical score, the score received from other vehicles, and the score assigned by the watchdog system.
Thus, the watchdog system performs appropriate management based on the reputation score of vehicle.
Li et al. \cite{li2012lie} applied a rate limiting mechanism to prevent flooding attacks.
The proposed mechanism limits the number of message creations and replications within a predefined time interval.
The rate limiting is achieved by adding a rate limiting certificate to the data package, which is issued and proven the validity by a trusted organization.
This work is also able to detect false counts, claimed by the attacker through the idea of claim-carry-and-check.
It can indeed effectively alleviate flooding attacks, however with certain limitations to defend against DDoS.

\subsubsection{Detection Combined with Sensor}
When the few vehicles do not collect sufficient messages for detection, it is difficult to judge the forged data, that is, it is difficult to judge from the data content level alone.
Yao et al. \cite{yao2018multi} measured the received signal strength indication (RSSI) value, summarized and captured four characteristics about RSSI:
a) Channel quality changes with time in VANETs.
b) Considering complex reflection, refraction, diffraction, and multi-path effects caused by obstacles, the signal propagation model requires setting different parameters for different environments.
However, the surrounding environment of vehicle is difficult to dynamically perceive, therefore the signal propagation model is difficult to adjust with such change.
c) The RSSI time series of attacking vehicles and fake sybil vehicles has very similar patterns.
d) The RSSI of sybil vehicles frequently changes, while normal vehicles do not.
In view of these characteristics, authors adopted time series similarity measurement and time series change point detection to detect sybil vehicles.
The proposed detection method does not rely on the trust relationship between vehicles, and instead considers the physical characteristics of vehicles that can not be hidden.
However, these characteristics may be exploited by attackers to hide their misbehavior.
Another disadvantage is that RSSI is greatly affected by the environment.

\subsection{Intrusion Detection System}
\subsubsection{IDS in V2G Network}
The smart grid is highly vulnerable to DoS attacks due to blackouts and disasters. 
DoS attacks attempt to delay, block, or disrupt communications, which can seriously degrade network performance. 
The deep learning-based IDS can promote the detection and mitigation of DoS attacks. 
It is deployed at the edge of smart grid to analyze information flow characteristics (such as time and frequency). 
The normal behavior pattern of information flow is constructed based on spatio-temporal features and context relations. 
Besides, the real-time schedulability analysis extracts timing specifications and model parameters of information flow. 
Then the IDS generates a real-time model that accurately describes the time characteristics, to represent the normal behavior of smart grid infrastructures. 
Finally, with the packet attribute analysis and the deep learning-based data consistency model, the normal behavior pattern of physical information system and the source of cyber attack can be traced. Table \ref{IDS_con} compares related works in the IDS of V2G and connected vehicles.

\begin{table*}[!htbp]\small
	\caption{Related Works in the IDS of V2G and Connected Vehicles}\label{IDS_con}
	\centering
	
	\begin{tabularx}{\textwidth}{cXX}
		\toprule
		Scheme & Primitives & Merit and Demerit\\
		\midrule
		\cite{2016Intrusion, 2020Deep, 2018Addressing, navaz2013entropy, stabili2017detecting} &
		\begin{itemize}
			\setlength{\itemsep}{0pt}			
			\setlength{\parsep}{0pt}
			\setlength{\parskip}{0pt}
			\item Low false-positive rate
		\end{itemize} &
		\begin{itemize}
			\setlength{\itemsep}{0pt}			
			\setlength{\parsep}{0pt}
			\setlength{\parskip}{0pt}
			\item Update database signatures when new threats emerge.
			\item Maintain a large database of known automobile network threats.
		\end{itemize}\\

		\cite{8994200, 8767141, 2017OCPP, cho2016fingerprinting, stabili2017detecting, koyama2019anomaly, kang2016intrusion, seo2018gids, barletta2020intrusion} &
		\begin{itemize}
			\setlength{\itemsep}{0pt}			
			\setlength{\parsep}{0pt}
			\setlength{\parskip}{0pt}
			\item Identify and forecast attacks based on the training and learning process.
		\end{itemize} &
		\begin{itemize}
			\setlength{\itemsep}{0pt}			
			\setlength{\parsep}{0pt}
			\setlength{\parskip}{0pt}
			\item Low-size injected attacks are difficult to detect.
		\end{itemize}\\
		\bottomrule
	\end{tabularx}
\end{table*}
Paria Jokar et al. \cite{2016Intrusion} proposed an intrusion detection and prevention system for ZigBee-based home Local Area Network (LAN) in smart grid. 
The system adopts model-based intrusion detection mechanism and machine learn-based intrusion prevention system (IPS), to protect the network from various types of attacks. 
This detection module makes a judgment of network state by extracting network features. 
In order to detect DoS attacks, Basnet \cite{2020Deep} implemented DNN and long short-term memory (LSTM) algorithms in charging piles to detect DoS attacks. 
In literature \cite{2018Addressing}, authors suggested adding encapsulation in the OCPP to avoid MITM attacks. 
Rubio et al. proposed a method that preserves privacy when sending smart meter data in OCPP. Besides, by studying possible security mechanisms and further add these to the EV infrastructure for upgrades or versions of the OCPP specification.

In literature \cite{8994200} and \cite{8767141}, the influence of tampering OCPP message on smart grid performance is shown. 
In literature \cite{8994200}, a security assessment of EV charging infrastructure was proposed, by classifing the cyber threats in the context of both public and private charging facilities. 
literature \cite{8767141} proposed a cyber-physical system approach to understand the interaction of various components within smart charging equipment. Furthermore, this work introduces the different types of vulnerabilities, attacks and approaches to improve CPS security.
Morosan and Pop \cite{2017OCPP} proposed a back-propagation neural network that is only applicable to detect DDoS attacks. 
Here, the back-propagation neural network is based on anomaly detection to classify the traffic into two classes: malicious traffic and benign. The difference between the two classes is the value from the output that delimits the classes using a threshold.

\subsubsection{IDS for In-vehicle Network}
The IDS attracts much attention because of its useability to quickly identify attacks.
In general, if there is any abnormality in the system, the network or host IDS will detect it and give an alarm.
According to the technical basis of intrusion detection, the IDS is divided into two categories:

\textbf{Signature-based IDS:} This IDS analyzes known attacks to extract their distinguishing features and patterns, which is called the signature. 
Signature-based IDS has the advantage of high detection rate for known attacks, but the disadvantage is that it is not able to detect unknown or new attacks.
Despite it having a high true-positive rate, when new threats emerge, it has to update its signature database \cite{navaz2013entropy}.
Signature-based IDSs have to maintain a large database of known automobile network threats \cite{stabili2017detecting}.
Extracting the attack characteristics of moving targets in real time is difficult and time-consuming.

\textbf{Anomaly-based IDS:} This type of IDS is also known as behavior-based IDS, by using supervised or unsupervised learning methods to build models based on features. 
The model can identify normal and abnormal network flow patterns, with advanced capability to detect unknown and new attacks.
This strategy is implemented using statistical and ML techniques.

\vspace{1mm}
\centerline{\textbf{[Statistical Approaches]}}
\vspace{1mm}

Cho et al. \cite{cho2016fingerprinting} created a clock-based IDS, which fingerprints each ECU based on the interval between data exchanges.
Authors applied the least square cost function and a sequential analysis technique called the cumulative sum algorithm to find anomalies.
It is difficult to detect low-volume injection data and counterfeit disabled ECU, just like prior techniques.

In addition, the Hamming distance is also used in intrusion detection in many studies.
For example, Stabili et al. \cite{stabili2017detecting} examined the payload of CAN and captured every bit in the data field.
If the calculated Hamming distance function obviously deviates from the expected Hamming distance function, the attack is identified.
Regarding the periodic CAN ID, Koyama et al. \cite{koyama2019anomaly} applied the quantized interval anomaly detection system, and checked the difference of CAN payload values through the quantized interval anomaly detection system.
It is proved that the solution successfully resists message injection attacks.
Instead, low-size injection assaults are difficult to be detected with their approach.

\vspace{1mm}
\centerline{\textbf{[ML Approaches]}}
\vspace{1mm}
The ML is applied in IDS to extract and learn normal/abnormal behavior, then create a model to identify attacks. 
ML-IDS is widely applied to process a large number of CAN traffic data because of its many characteristics.
By using DNN, the rule pattern is learned from unstructured samples, and the deviation is identified as abnormal \cite{kang2016intrusion}. Meanwhile, the unsupervised deep belief network (DBN) is adopted to pre-process the data and find the regular trend.
In addition, for the dataset, authors simulated CAN frames by running network software, and added noise to provide randomness.
This solution processes the original CAN bus data without data reduction and extraction in the pre-processing stage. 
Seo et al. \cite{seo2018gids} applied generative antagonism network to detect patterns in CAN data without classification. The scheme is tested in DoS, frame fuzzing and spoofing attacks.
Approaches based on the decision tree are able to classify normal or anomalous CAN data.
During the training step, decision trees require supervised labeled data sets to make subsequent judgments.
By using a gradient boosting approach, Barletta et al. \cite{barletta2020intrusion} applied a regression decision tree to build a classifier.
For designing the selection method, authors applied entropy to evaluate CAN ID and the data payload time.
For the dataset, authors used $750000$ real records of CAN data, and made changes to the dataset in the form of random value for testing.

\subsection{Firewall}
A firewall is a barrier between public and private networks. 
It can detect attacks and filter malicious network flow. 
The new generation firewall can also combine DoS attack detection and network protocol identification technology. 
The former is used to filter attack flow and reduce the threat of attack, while the latter is used to deal with injection and spoofing. Table \ref{firewall} lists related works on firewall.
Here, transport layer security (TLS) and secure sockets layer (SSL) protocols can be combined with SHA, HMAC and other encryption technologies to complete the operation of communication to data check \cite{MaHongzhe}.
\begin{table*}[!htbp]\small
	\caption{Related Works on Firewall}\label{firewall}
	\centering
	
	\begin{tabularx}{\textwidth}{cXX}
		\toprule
		Scheme & Primitives & Merit and Demerit\\
		\midrule
		\cite{khosroshahi2016security} &
		\begin{itemize}
			\setlength{\itemsep}{0pt}			
			\setlength{\parsep}{0pt}
			\setlength{\parskip}{0pt}
			\item The firewall was divided into three parts.
		\end{itemize} &
		\begin{itemize}
			\setlength{\itemsep}{0pt}			
			\setlength{\parsep}{0pt}
			\setlength{\parskip}{0pt}
			\item Analyze packets by header, port number, service, etc.
		\end{itemize}\\

		\cite{diovu2017cloud} &
		\begin{itemize}
			\setlength{\itemsep}{0pt}			
			\setlength{\parsep}{0pt}
			\setlength{\parskip}{0pt}
			\item A cloud-based firewall
		\end{itemize} &
		\begin{itemize}
			\setlength{\itemsep}{0pt}			
			\setlength{\parsep}{0pt}
			\setlength{\parskip}{0pt}
			\item Combine cloud computing technology with traditional infrastructure security capabilities.
		\end{itemize}\\

            \cite{liu2019isrf} &
		\begin{itemize}
			\setlength{\itemsep}{0pt}			
			\setlength{\parsep}{0pt}
			\setlength{\parskip}{0pt}
			\item A fog-based firewall (a semantic reasoning scheme)
		\end{itemize} &
		\begin{itemize}
			\setlength{\itemsep}{0pt}			
			\setlength{\parsep}{0pt}
			\setlength{\parskip}{0pt}
			\item Utilize a context selection based on semantics to defend against threats.
		\end{itemize}\\
		\bottomrule
	\end{tabularx}
\end{table*}
\subsubsection{Firewall in Smart Grid}
As part of the smart grid system, the host and server's IP address in the network can be determined. 
Therefore, the packet filtering function of firewall can be achieved with a whitelist. 
First of all, the firewall checks network flow based on the protocol type, port number, and destination IP address, so as to judge whether traffic complies with specified standards, regulations and whitelists. 
Then, the location of firewall can be set according to the structure of smart grid network. 
Note that, the firewall needs to facilitate the management of relevant staff. 
Finally, the firewall in smart grid also needs to establish a secure access policy to identify and filter access with different permissions \cite{alcaraz2016policy}. 
For example, the smart grid management center can access substations, but the substation can not access the core network of the control center. 
The firewall of smart grid in \cite{khosroshahi2016security} was divided into three layers. The first layer of firewall is based on analyzing the IP packet and its header fields. 
The second layer of firewall analyzes the packet based on the port number of origin and destination process, control codes, number field, and acknowledgment field. The third layer protects the smart grid based on the type of service and application. 
Diovu et al. \cite{diovu2017cloud} proposed a cloud-based firewall to mitigate DDoS attacks in a smart grid advanced metering infrastructure (AMI) network. 
It is able to detect and mitigate the effects of DDoS attacks, additionally prevent them before attacks are launched. 
It combines the latest technological advances in cloud computing technology with traditional infrastructure security capabilities.

\subsubsection{Firewall in EV}
Firewalls on the EV side are also being developed to deal with a variety of attacks. 
One of the characteristics of in-vehicle network is that network configuration information is essentially static. 
The original equipment manufacturer (OEM) typically has a communication matrix or database that defines the rules for communication between devices in the vehicle \cite{ELREWINI2020100214}. 
As a result, the firewall technology of in-vehicle network is generally based on "whitelist" filtering rules. 
In other words, according to the predefined communication matrix, the data that meets the conditions will be forwarded or processed, otherwise it will be discarded. 
In \cite{liu2019isrf}, a fog-based firewall system for internet of vehicle (IoV) is proposed to protect the security of IoV. 
Through mining potential threats in packets, a semantic reasoning scheme is proposed, by utilizing a context selection scheme based on semantics. This fog-based firewall system can defend against both content threats and network layer threats.

Note that, In-vehicle firewalls are divided into interdomain and boundary firewalls. 
Interdomain firewalls control the forwarding of interdomain traffic. 
Interdomain security policies match traffic by IP address, time range, service (port or protocol type), user, etc., and implement packet filtering control (permit or deny) for CAN packets that meet the requirements. 
The interdomain firewall is used to control the communication between the external devices. 
It matches the network flow by IP address, time range, service (port or protocol type), etc., and controls packet filtering (permit or deny) for the CAN packets that meet the conditions, allowing or denying the communication between the external devices and the internal devices.

Boundary firewalls are "gatekeepers" that use one or more of the following methods to control network flow. 
First, the static packet filtering is based on the header information of data. 
Second, stateful checking is a common firewall method that records outgoing network flow, and allows network flow that corresponds to the initial request to be sent back. 
Third, vehicles utilize the vehicle gateway to manage communication between the in-vehicle network and external systems. 
Similar to the home gateway, the vehicle gateway isolates the internal ECU from attack, which has become the vulnerability for hackers. 
Therefore, vehicle manufacturers and OEMs need to protect the vehicle gateway from attacks. 
Meanwhile, there is also necessity to implement firewall filtering rules to control network flow, which is forwarded to the internal network.

\section{Network Application Security}\label{networkapp}
Network application security is a challenge, which involves the complexity proportional to the environment. 
Protecting subnets, switches, routers, and firewalls is a traditional field, but it is difficult to improve security by using EC, SDN and other new services. 
Meanwhile, in EC, SDN, and other services, security risks are still prominent in all network environments.

\subsection{Software-Defined Networking Security}
In 5G network, the SDN architecture aims to achieve flexible control and efficient data forwarding.
Due to various control functions (routing decision-making function, etc.) in the network being separated, unified scheduling and distribution network is realized by a central controller. 
The architecture of a SDN includes a data layer and application layer. 
The data layer is responsible for formulating network protocols, integrating and processing network data, thus providing network services. 
The application layer is responsible for providing application services.

Using software to define network architecture is beneficial to make the smart grid architecture flexible and reliable. 
The application of SDN technology in the smart grid includes the data layer responsible for processing communication network data and forwarding power network data. 
The control layer is responsible for providing communication network and smart grid integration services. 
The application layer is responsible for providing application services for smart grid.
\begin{table*}[!htbp]\small
	\caption{The Applications of SDN in the Security of Smart Grid and IoV}\label{SDNSmartG}
	\centering
	\begin{tabularx}{\textwidth}{cXX}
		\toprule
		Application Field & Application Scenario & References\\
		\midrule
		Smart Grid &
		\begin{itemize}
			\setlength{\itemsep}{0pt}			
			\setlength{\parsep}{0pt}
			\setlength{\parskip}{0pt}
			\item Monitor network traffic
			\item IDS/IPS and firewall
			\item SCADA
		\end{itemize} &
		\begin{itemize}
			\setlength{\itemsep}{0pt}			
			\setlength{\parsep}{0pt}
			\setlength{\parskip}{0pt}
			\item \cite{8567005}
			\item \cite{8251983}
			\item \cite{da2016one}
		\end{itemize}\\	
		IoV &
		\begin{itemize}
			\setlength{\itemsep}{0pt}			
			\setlength{\parsep}{0pt}
			\setlength{\parskip}{0pt}
			\item Service interruption
			\item VANETs
		\end{itemize} &
		\begin{itemize}
			\setlength{\itemsep}{0pt}			
			\setlength{\parsep}{0pt}
			\setlength{\parskip}{0pt}
			\item \cite{s16122077}
		\end{itemize}\\
		\bottomrule
	\end{tabularx}
\end{table*}

Table \ref{SDNSmartG} shows the applications of SDN in the security of smart grid and IoV. Sedef et al. \cite{8567005} detected and mitigated insider attacks in smart grid, by monitoring and analyzing network traffic in real time using SDN concepts. 
Due to the programmability of SDN, the development of cyber security devices such as firewalls and IDS/IPS are promoted. 
Duha et al. \cite{8251983} further studied smart grid SDN integration resilience against attacks. 
It dynamically configures the network by using SDN to prevent attacks and isolate them if possible. 
In \cite{da2016one}, authors proposed an IDS that leverages the properties of SDN and supervisory control and data acquisition (SCADA) for traffic classification.

The development of SDN technology also enhances the security in IoV. First of all, intelligence and dynamic programmability of SDN can solve the problem of service interruption, caused by transmission interference during the communication process between EVs and RSUs. 
Secondly, in queuing scenarios (await for traffic lights), the application of SDN in IoV can mitigate replay and jamming attacks \cite{s16122077}. 

\subsection{Edge Computing Security}
\begin{table*}[!htbp]\small
	\caption{The Applications of EC in the Security of Smart Grid and IoV}\label{ECSmartG}
	\centering
	\begin{tabularx}{\textwidth}{cXX}
		\toprule
		Application Field & Application Scenario & References\\
		\midrule
		Smart Grid &
		\begin{itemize}
			\setlength{\itemsep}{0pt}			
			\setlength{\parsep}{0pt}
			\setlength{\parskip}{0pt}
			\item UAV monitors transmission lines
			\item Detect NTL
			\item Authentication
            \item CASB
		\end{itemize} &
		\begin{itemize}
			\setlength{\itemsep}{0pt}			
			\setlength{\parsep}{0pt}
			\setlength{\parskip}{0pt}
			\item \cite{9532792}
			\item \cite{2020Edge}
			\item \cite{2021LightweightChen}
                \item \cite{ahmad2022rsm}
		\end{itemize}\\	
		IoV &
		\begin{itemize}
			\setlength{\itemsep}{0pt}			
			\setlength{\parsep}{0pt}
			\setlength{\parskip}{0pt}
			\item VANETs
			\item Verification delay
		\end{itemize} &
		\begin{itemize}
			\setlength{\itemsep}{0pt}			
			\setlength{\parsep}{0pt}
			\setlength{\parskip}{0pt}
			\item \cite{2017Meet}
			\item \cite{Arwa2018An}
		\end{itemize}\\
		\bottomrule
	\end{tabularx}
\end{table*}

The EC decentralizes the processing of data, operation of applications and realization of functional services from the central server to the edge of network \cite{2019Raza}. 
Since EC is close to users, it is expected to meet the requirements of high bandwidth and low latency to improve service quality. 
Meanwhile, by deploying services and caches at the edge of network, the central network can not only reduce network congestion but also efficiently respond to user requests. 
For example, in the field of unmanned driving, unmanned vehicles need to respond quickly to the surrounding infrastructures in a high-speed moving state. 
In this case, the response time becomes crucial. 
To achieve this efficiency, it is necessary to rely on EC.

In recent years, vehicular edge computing (VEC) has been brought into vehicular networks to enhance their computing capabilities \cite{2019Raza}. 
Through VEC technology, service providers directly provide hosting services, reducing latency and improving the quality of service. 
The VEC targets applications with widely distributed deployments. 
Since VEC further extends the benefits brought by cloud computing services to the network edge \cite{7498684}, the goal of VEC is to apply widely distributed applications.

Along with EC mechanism, the deployment location of EC nodes is shifting away from central security protection.
Nevertheless, the security level of EC nodes will be reduced, making it easier for attackers to access the hardware of EC nodes. 
Attackers can access network ports through illegal connections to obtain data transmitted over the network. In addition, traditional cyber attack methods still threaten EC systems, such as malicious code intrusion, buffer overflow, data theft, tampering, loss, falsification of data, etc.

To reduce the bandwidth and storage requirements of a cloud, Huang et al. \cite{2017Meet} proposed a model named Meet-Fog. 
The Meet-Fog is based on fog and Meet-Table technologies to accurately spread negative messages like certificate revocation list (CRL) in VANETs. 
This scheme completely transfers the computing requirements to the edge instead of cloud.

In \cite{Arwa2018An}, authors proposed an efficient revocation architecture for VANETs. 
This scheme increases the effectiveness of certificate status checking process by utilizing EC and Merkle hash tree. 
Compared with other methods using CRL check, the verification delay is also reduced. 
To protect edge nodes from accessing the cloud, Ahmad et al. \cite{ahmad2022rsm} introduced an overview of existing cloud access security broker (CASB) strategies and how strategies are applied to the cloud computing environment. 
Authors applied principal component analysis (PCA) to study different independent parameters that affect CASB. 
The result of analysis is to determine five parameters that affect PCA analysis.

Applying EC technology to smart grid enables plug-and-play and interoperability \cite{9532792}. For example, when an unmanned aerial vehicle (UAV) monitors power lines, the EC can be integrated with system services. 
The connection between the microgrid control center and EC nodes, is helpful in solving the real-time demand-side power consumption scenario.

In \cite{2020Edge}, authors proposed an EC-enabled non-technical loss (NTL) fraud detection scheme called ENFD. It is supported by an EC framework, which provides a big data security analytic (BDSA) algorithm to analyze big data to detect NTL fraud in the smart grid. 
A lightweight key exchange and mutual authentication protocol \cite{2021LightweightChen} was proposed for edge-based smart grid environments. 
The design of this protocol applies a one-way hash function, XOR computation, and ECC instead of another heavy cryptographic function. 
Table \ref{ECSmartG} shows the applications of EC in the security of smart grid and IoV.

\section{Future Research Directions}\label{future}
In previous sections, perspectives from data security, network management and network application security have been reviewed in detail.
Here, it is clear that most of literature have paid attention on data security as that is the dominant element in low-carbon transportation, is of importance to trigger service in safety manner.
Besides, the network management security is complementary to data security, based on the ground that the secure data can be utilized as guidance for enabling network management, including monitoring and detecting malicious network devices (normally those have passed the authentication or is eligible to participate data encryption/decryption).
Thanks to this, both the defense technologies in SDN and EC, will be reliable to resist against threats from external attackers (defensed through authentication and encryption technologies) and those malicious attackers (detected through trust management, malicious behavior detection, IDS and firewall).

Indeed, there are still open issues remaining to communities. The primary direction is to promote the intelligent level of vehicles in low-carbon transport system, so as to benefit from much powerful digitalized services and cyber security technologies. 
Other emerging directions such as detachable security technology, zero knowledge proof are worthwhile investigation in line with the integrated control for low-carbon transportation. These potential directions are introduced as follow:

\subsection{Intelligent Connected Vehicles (ICV)}
The ICV is widely integrated with vehicle information and communication, artificial intelligence, software, Internet, etc.
Meanwhile, the ICV requires coordinated development across industries and fields.
At present, the ICV is in the stage of rapid increase with penetration rate.
The ICV can integrate millions of mobile applications into EVs, and achieve a breakthrough in EV applications.
Thus, the ICV are expected to become a new generation of super terminals after the smartphone.
In addition, due to the rampant cyber attacks, the security of ICV needs to be concerned.

\subsection{Detachable Security Technology}
As the continuous improvement of various technologies and the emergence of new technologies, from hardware architecture to software algorithms, each technology is combined with each other to achieve dedicated functions. 
However, these technologies are not combined with certain level, and also need to achieve a balance between function and performance.

The problem of multi-functional and simple firewalls is a typical example. 
For choosing a firewall, its performance and functions should be considered. 
The more functions a simple firewall reduces, the higher its performance will be. 
If the function of a multi-functional firewall increases, its performance will be reduced accordingly. 
Therefore, both the multi-functional firewall and the simple firewall need to strike a balance between function, performance, applicability and stability.

\subsection{Integrated Control Problem}
In the context of policy support and environmental protection, people will be more inclined to buy EVs.
Along with the number of EVs rapidly increasing, the communication in the IoV needs attention.
Then, when a large number of EVs are connected to the smart grid, the impact on the smart grid system will become greater.
Therefore, the integrated control problem of EV smart charging and smart grid collaborative technology is also one of the future research directions. It can be mainly divided into the identity authentication of the EV with charging demands, the secure communication in IoV, and the stable operation of smart grid system. 
How to ensure secure access, communication and control the charging/discharging time is particularly important.

\subsection{Zero Knowledge Proof}
The security of intelligent transportation relies on large, heterogeneous, and multi-source data. To ensure computational security, zero knowledge proof is gradually applied to intelligent transportation systems \cite{2020Privacyzero}. 
Zero knowledge proof \cite{1992On} is to make the verifier believe that the prover knows the proof without disclosing private information. Currently, zero knowledge proof is applied to the privacy protection of convolutional neural network model parameters, such as the number of convolutional layers and weights. 
That is to say, under the condition that the verifier can not obtain model parameters, zero knowledge proof is applied to ensure that the training results are trustworthy and tamper-proof.

However, zero knowledge proof still faces a large amount of computation and does not support scenarios with limited resource. 
Additionally, the system must generate public parameters for the prover and verifier in advance to complete the verification. 
In the future, zero knowledge proof without bilinear mappings should be focused on.
Moreover, on the basis of not affecting the availability of data, zero knowledge proof is applied to improve the privacy of data.

\subsection{Promotion of Global-Level Policy}
Given the early stage development of low-carbon transportation, there is still necessary to design the top-level policy and accelerate technology innovation.
Firstly, the low-carbon manufacturing is realized based on the green energy supply, and sequentially the energy consumption of EVs is promoted through clean energy.
In the light of this, it is necessary to continuously optimize the energy structure of urban transportation, and increase the proportion of green energy source.
Secondly, accelerating the application of emerging technologies such as vehicle-road coordination, 5G is with potential to realize low cost transportation.
Thirdly, strengthening the cooperation between EVs and charging infrastructures realizes full and timely information interaction between EVs and roadside infrastructures.
Therefore, making full use of the emerging information technologies (such as the 5G, big data, artificial intelligence, SDN, EC and blockchain) are indeed to promote the integration of EVs and transportation industries.
Finally, as several elements from interdisciplinary areas are included, the low-carbon transportation system is vulnerable to known/unknown attacks. Aiming to guarantee the security of low-carbon transportation, it is of great importance to well deploy security defense systems and establish IDS.

\section{Concluding Remarks}
Concerning the increasing attention on achieving low-carbon transportation and necessity to secure the system through cyber security technologies, this review firstly introduces the concept and background of low-carbon transportation. 
Then, based on identifying typical attacks within the ecosystem of low-carbon transportation system, this review classifies and reviews emerging defense technologies from the aspects of data security, network management security and network application security, covering up-to-date technical advance that have been contributing to communities.
Finally, along with existing contribution, this review further highlighted several future directions that cover the cyber secure low-carbon transportation system from evolution of vehicles, compatibility of defense technologies integration and potential impact on unlocking the cyber security and system reliability.

The analysis of literature is objective summary and classification into different technical taxonomy, therefore there is no issue about sources of error in analysis as all references are published at high quality international journals and conferences. As the angle of review is from cyber security, the purpose of this survey is to attract attention from both academic and industrial communities, thus persistent efforts in promoting the literature are required.


\bibliographystyle{elsarticle-num}
\bibliography{cas-refs}

\end{document}